\documentclass[11pt,dvips]{article}
\usepackage{amsfonts}
\usepackage{amsmath}
\usepackage{amssymb}
\usepackage{graphicx}
\usepackage{epsfig}
\usepackage{multirow}
\usepackage{dcolumn}
\usepackage{bm}
\usepackage{cite}
\usepackage[small]{caption}
\usepackage[usenames]{color}
\usepackage{enumerate}
\usepackage{hyperref} 
\usepackage{breakurl}

\def\wt{\widetilde}
\def\sla#1{\setbox0=\hbox{$#1$}\dimen0=\wd0
      \setbox1=\hbox{/} \dimen1=\wd1 \ifdim\dimen0>\dimen1
      \rlap{\hbox to \dimen0{\hfil/\hfil}} #1
      \else
      \rlap{\hbox to \dimen1{\hfil$#1$\hfil}}
      /   \fi}
\def\MET{\sla{E}_{\rm T}}
\def\MPT{\sla{p}_{\rm T}}

\def\lsim{\raise0.3ex\hbox{$\;<$\kern-0.75em\raise-1.1ex\hbox{$\sim\;$}}}
\def\gsim{\raise0.3ex\hbox{$\;>$\kern-0.75em\raise-1.1ex\hbox{$\sim\;$}}}
\topmargin
-1.5cm
\textwidth
15.05cm
\textheight
20.5cm
\oddsidemargin
0.7cm
\evensidemargin
0.7cm

\def\Sn1{{S^0}^*}
\def\Pn1{{P^0}^*}
\def\Cn1{{S^\pm}^*}

\def    \beq            {\begin{equation}}
\def    \eeq            {\end{equation}}
\def    \bea           {\begin{eqnarray}}
\def    \eea           {\end{eqnarray}}

\def\fbi{{\rm fb}^{-1}}
\def\la{\lambda}
\def\ka{\kappa}

\def \wt {\widetilde}
\def\h{h^0_}
\def\a{a^0_}

\def\bmumu{B^0_s\to\mu^+\mu^-}

\def\neut{\widetilde \chi^0}
\def\cha{\widetilde \chi^\pm}

\newcommand{\unit}{\leavevmode\hbox{\small1\kern-3.6pt\normalsize1}}

\allowdisplaybreaks

\def\MET{E_{\rm T} \hspace{-1.2em}/\;\;}
\def\MPT{p_{\rm T} \hspace{-1.2em}/\;\;}

\begin{document}

\thispagestyle{empty}
\begin{flushright}
  FTUAM-13-21\\
  IFT-UAM/CSIC-13-083\\
  CERN-PH-TH/2013-176\\

\end{flushright}

\begin{center}
  {\bf {\LARGE Collider signatures of a light NMSSM pseudoscalar \\
      \vskip 0.1in
      in neutralino decays in the light of LHC results}}
\end{center}

\renewcommand*{\thefootnote}{\fnsymbol{footnote}}
\begin{center}
  {\large
    David~G.~Cerde\~no$^{a,b,}$\footnote{Email: davidg.cerdeno@uam.es},
    Pradipta~Ghosh$^{a,b,}$\footnote{Email: pradipta.ghosh@uam.es},
    Chan~Beom~Park$^{c,}$\footnote{Email: chanbeom.park@cern.ch}, and
    Miguel Peir\'o$^{a,b,}$}\footnote{Email: miguel.peiro@uam.es (MultiDark Scholar)}
\end{center}

\begin{center}
$^a${\em Departamento de F\'{\i}sica Te\'{o}rica, Universidad
Aut\'{o}noma de Madrid,\newline
Cantoblanco, E-28049 Madrid, Spain}\\
$^b${\em Instituto de F\'{\i}sica Te\'{o}rica UAM/CSIC, Universidad
Aut\'{o}noma de Madrid,\newline
Cantoblanco, E-28049 Madrid, Spain}\\
$^c${\em TH Division, Physics Department, CERN, CH-1211 Geneva 23,
Switzerland}
\end{center}

\begin{center}
  \begin{abstract}
    \noindent
    We investigate signatures induced by a very light pseudoscalar Higgs in
    neutralino decays in the Next-to-Minimal Supersymmetric Standard
    Model (NMSSM) and determine their observability at the LHC.
    We concentrate on scenarios which feature two light scalar Higgs bosons
    (one of them is SM-like with a mass of 125 GeV and a singlet-like
    lighter one) with a very light (singlet-like) pseudoscalar Higgs in the
    mass range $2m_\tau<m_{\a1}<2m_b$.
    We consider neutralino-chargino pair production and the subsequent
    decay $\neut_{2,3}\to\neut_1\a1$, which leads to topologies
    involving multi-leptons and missing transverse energy.
    We determine a set of selection cuts that can effectively isolate
    the signal from backgrounds of the Standard Model or the Minimal
    Supersymmetric Standard Model. We also exemplify the procedure with a
    set of benchmark points, for which we compute the expected number
    of events and signal strength for LHC with 8 TeV center of mass
    energy. We show that this signal can already be probed for some
    points in the NMSSM parameter space.
  \end{abstract}
\end{center}

\newpage
\section{Introduction}
\setcounter{footnote}{0}
\renewcommand*{\thefootnote}{\arabic{footnote}}

The Next-to-Minimal Supersymmetric Standard Model (NMSSM) (see, e.g.,
Ref.\cite{Ellwanger:2009dp} for a review) is a well
motivated extension of the MSSM. In the NMSSM a new singlet superfield $\hat S$
is included in order to provide a dynamical mechanism by which
the Higgsino mass parameter, $\mu$, is naturally of the order of the
electroweak (EW) scale, thereby addressing the so-called ``$\mu$ problem''
\cite{Kim:1983dt}.
The NMSSM leads to a very interesting Higgs phenomenology, due to the
presence of an extra scalar Higgs and a pseudoscalar Higgs. These
states can be very light without violating current collider
constraints, provided that they are mostly singlet-like.
Moreover,
within the NMSSM a new contribution to the tree level Higgs mass
\cite{Drees:1988fc,Ellis:1988er,Binetruy:1991mk,Espinosa:1991gr,
Espinosa:1992hp}, coming from the
$\lambda \hat S \hat H_u \hat H_d$ term in the superpotential, makes it easier to
obtain a relatively heavy Standard Model (SM) like Higgs boson
while reducing somehow the fine-tuning \cite{BasteroGil:2000bw,
King:1995vk,Dermisek:2005ar,Dermisek:2007yt,Ellwanger:2011mu,
Arvanitaki:2011ck,Kang:2012sy,Cheng:2012pe,Perelstein:2012qg,
Agashe:2012zq,King:2012tr}.
This is favoured by the LHC observation of a Higgs boson with a mass
in the $2\,\sigma$ range $124-126.8$~GeV ($124.5-126.9$~GeV)
by ATLAS (CMS)~\cite{Higgs,CMS:2012gu,ATLAS:2013mma,CMS:yva}.
It has also been shown that in some regions of the parameter 
space another light, singlet-like, Higgs $(h^0_1)$ can also be present,
with a mass around $98$~GeV \cite{Belanger:2012tt,Aparicio:2012vk,CGP1,
Kang:2013rj,Bhattacherjee:2013vga}, motivated by the
small excess in the LEP search for $e^+e^-\to Z h$, $h\to b \bar{b}$
\cite{Barate:2003sz,Schael:2006cr,Drees:2012fb}, or even lighter
\cite{Ellwanger:2003jt,Ellwanger:2004gz,Ellwanger:2005uu,
Djouadi:2008uw,Stal:2011cz,CGP1}.

Some of these scenarios also present a very light pseudoscalar Higgs
boson, $a_1^0$, \cite{Ellwanger:2003jt,Ellwanger:2004gz,Ellwanger:2005uu,
Djouadi:2008uw,CGP1,Bhattacherjee:2013vga}, which is very
appealing from the point of view of LHC signatures.
For example, in Ref.\,\cite{CGP1} we investigated how a very light
pseudoscalar in the mass range $2 m_\tau< m_{a_1^0}<2m_b$ could be
probed in multilepton decays of the scalar Higgses, $h_{1,\,2}^0 \to
a_1^0 a_1^0 \to 4\ell + \MET$ (where $\ell =e$, $\mu$, and
hadronically-decaying $\tau$, $\tau_h$ and 
$\MET$ denotes the missing transverse energy), and showed how the pseudoscalar
mass can be reconstructed.
Another remarkable effect of a light pseudoscalar is that it can be
copiously produced in the decays of neutralinos
$\neut_{2,3}\to\neut_1a_1^0$, constituting a characteristic NMSSM
signature~\cite{Franke:1995tf,Choi:2004zx,Cheung:2008rh,Stal:2011cz,
Das:2012rr}.
It should be noted that these scenarios are extremely sensitive to the
Higgs properties and therefore very affected by the recent
experimental constraints. For example, the latest measurements
of the SM-like Higgs ($h_{\rm SM}$) \cite{ATLAS:2013mma,CMS:yva} forbid a
sizable contribution to non-SM decay modes, such as
$h_{\rm SM}\to a_1^0 a_1^0$, $h_{\rm SM}\to \h1 \h1/h^{0^*}_1$ 
or $h_{\rm SM}\to\neut_1\neut_1$, and
therefore motivates the reanalysis of the phenomenology associated to
neutralino decays.

In this work we  concentrate on scenarios which feature two light
scalar Higgses (one of them SM-like with a mass of approximately 125
GeV and a singlet-like lighter one) with a very light singlet-like
pseudoscalar (in the mass range $2m_\tau<m_{\a1}<2m_b$).
We carry out a systematic search for regions of the NMSSM parameter
space in which the branching ratio BR$(\neut_{2,3}\to\neut_1a_1^0)$ is
sizable.
For the range of masses considered, the pseudoscalar predominantly
decays into a pair of taus, $\a1\to \tau^+\tau^-$, leading to an
abundance of leptons in the final state.
Therefore, the resulting LHC phenomenology features multi-lepton
signals with missing transverse energy in the decay chains after
neutralino/chargino pair production,
$\neut_{2,3}\cha_1\to \ell^+\ell^-\ell^\pm+\MET$, $\neut_{3}\cha_1\to
2\ell^+2\ell^-\ell^\pm+\MET$, and
$\neut_i\neut_j\to n(\ell^+\ell^-)+\MET$, with $n=2,3,4$ and $i,j=2,3$.

Multi-lepton final states have always been considered an important
probe for supersymmetry (SUSY) searches in colliders
\cite{Nath:1987sw,Barbieri:1991vk,Lopez:1992ss,Baer:1993tr,
Lopez:1994dm,Mrenna:1995ax,Frank:1995dj,Baer:1996ms,
Abbott:1997je,Barger:1998wn,Barger:1998hp,Matchev:1999nb,
Baer:1999bq,Matchev:1999yn,Bisset:2003ix,Abazov:2005ku,
Aaltonen:2007af,Sullivan:2008ki,Aaltonen:2008pv,Abazov:2009zi,
Bhattacharyya:2009cc,Mondal:2012jv,tri-ATLAS,Aad:2012hba,
Chatrchyan:2012pka,Aad:2012xsa,CMS:aro,
ATLAS:2012uks,ATLAS:2013rla,Chatrchyan:2012mea,CMS:oxa}.
The unusual source of leptons ($\a1\to \tau^+\tau^-$) considered in this work 
within the context of the NMSSM with a pair of light
scalar Higgses
requires the modification of conventional search strategies.
In this paper, we define a set of event selection
cuts that will allow the studied signal to be distinguished from the SM and
MSSM backgrounds.
All the recent experimental bounds on the Higgs sector are included,
as well as the constraints on the masses of supersymmetric particles and low-energy
observables. We also take into account bounds on the neutralino relic
abundance and on its elastic scattering cross-section off quarks.

The paper is organised as follows. In Sec.\,\ref{sec:benchmark},
we present the results of a scan in the NMSSM parameter space,
applying the most recent experimental
constraints on Higgs sector and low-energy observables
as well as constraints from dark matter searches. We also
determine the regions of interest for our analysis.  In Sec.\,\ref{sec:prod},
we describe the collider phenomenology of the signal and
define a set of effective event
selection cuts for background suppression. 
We further estimate the relevant backgrounds and the
signal significance of the resulting signatures for some selected
benchmark points. Finally, we present our concluding remarks
in Sec.\,\ref{sec:conclusions}.

\section{Light Higgs scenarios in the NMSSM and choice of benchmark points}
\label{sec:benchmark}

The $\mathbb{Z}_3$ invariant NMSSM superpotential 
(see e.g., \cite{Ellwanger:2009dp}) reads
\beq
W = W_{\rm MSSM}^\prime -\epsilon_{ab} \lambda \hat S \hat H^a_d\hat H^b_u
+\frac{1}{3}\kappa \hat S^3,
\label{superpotential}
\eeq
where $W_{\rm MSSM}^\prime$ is the MSSM superpotential \cite{Martin:1997ns} 
without the bilinear $\epsilon_{ab} \mu \hat H^a_d\hat H^b_u$ term, with $\epsilon_{12}=1$.
$\hat H_u, \, \hat H_d$ are two $\rm{SU(2)}$-doublet Higgs superfields
and $\hat S$ is a new superfield, singlet under the SM gauge group. The 
superpotential incorporates two new couplings, $\lambda$ and $\kappa$.
The Lagrangian contains new soft SUSY-breaking terms, which
include the trilinear parameters $A_\lambda$ and $A_\kappa$, and the soft
mass parameter for the singlet, $m_S$. After EW 
symmetry breaking takes place, the neutral components of the Higgs
fields $H_{u,d}$ and the singlet $S$ acquire non-vanishing vacuum
expectation values, $v_{u,d}$ and $v_s$, respectively. Consequently, an effective
term $\mu=\lambda v_s$ is generated which is naturally of the order
of the EW scale.
In terms of the field content, the singlet mixes with the doublet Higgs
states, giving rise to three CP-even $(\h1,\,\h2,\,\h3)$ and two
CP-odd $(\a1,\,\a1)$ states, whereas the singlino mixes with the
neutralinos, inducing a fifth eigenstate $(\neut_1, \,\dots
,\,\neut_5)$, with interesting implications for dark matter searches
(see, e.g., Refs.\,\cite{Cerdeno:2004xw,Belanger:2005kh,Gunion:2005rw,
Cerdeno:2007sn, Hugonie:2007vd,Aalseth:2008rx,Das:2010ww,Cao:2011re}).

We have carried out a scan with a reduced set of the NMSSM parameters
defined at the EW scale, which are detailed in Table\,\ref{parameters}.
The scan is performed with \textsc{nmssmtools
  3.2.1}~\cite{nmssmtools,Ellwanger:2005dv,Ellwanger:2006rn}, linked
with \textsc{MultiNest 2.9} \cite{Feroz:2007kg,Feroz:2008xx} to
explore the parameter space efficiently. 
We impose the grand unification relation for the gaugino soft masses, 
which implies $M_1 = 1/2 M_2 = 1/6 M_3$. 
Fixed values are used for the trilinear parameters,
$A_{t}=1800$~GeV, $A_{b}=1000$~GeV, and $A_{\tau}=-1600$~GeV, as well
as for the soft scalar masses of sleptons and squarks,
$M_{\wt{L}_i}=M_{\wt{e}_i^c}=300$ GeV and
$M_{\wt{Q}_i}=M_{\wt{u}_i^c}=M_{\wt{d}_i^c}=1000$~GeV, respectively, where
the index $i$ runs over the three families.

We consider the most recent experimental limits
on sparticle masses~\cite{susy-latest,CMS-susy-multijet,
ATLAS-susy13,CMS-susy13} derived for simplified
SUSY model. For 
first two generation of squarks we set an optimized
lower bound of $1$ TeV consistent with Refs.~\cite{ATLAS-susy13,CMS-susy13}.
In the same way, the lower limit on lightest stop mass is
set to be $m_{\tilde t_1}> 650$ GeV, while for the lightest sbottom
we impose $m_{\tilde b_1}> 700$ GeV \cite{ATLAS-susy13,CMS-susy13}.
Regarding the gluino mass, we consider
$m_{\tilde g}>1.2$ TeV, which holds when the lightest supersymmetric 
particle (LSP) has non-vanishing mass,
and can be independent of squark masses \cite{ATLAS-susy13}. 
The situation is far more complicated with EW neutralinos
and charginos depending on modes of decay and $m_{\rm LSP}$. 
Thus, for lighter chargino we stick to the LEP lower limit of $94$ 
GeV \cite{Beringer:1900zz}.  The lightest neutralino, if light enough, is constrained by 
its contributions to the invisible $Z$ and Higgs bosons decays.
We also take into account experimental bounds on low energy 
observables~\cite{Asner:2010qj,Chatrchyan:2012rga,Aaij:2012ac,
Aad:2012pn, Lees:2012ju,Aaij:2012nna,Aaij:2013aka,Chatrchyan:2013bka,Galanti-talk,
Amhis:2012bh,Bennett:2006fi,Gray:2010fp,Jegerlehner:2009ry,
Davier:2010nc, Hagiwara:2011af} together with cosmological constraints
on the dark matter abundance \cite{Komatsu:2010fb,Ade:2013zuv} and
limits on its spin-independent elastic cross section with quarks from
direct detection experiments
\cite{Angle:2011th,Aprile:2012nq,Ahmed:2009zw,Ahmed:2011gh,Agnese:2013rvf}.
Regarding the Higgs sector we impose the presence of a SM-like Higgs,
which in our case corresponds to the second lightest mass state,
in the range $123$~GeV$\leq m_{\h2}\leq 127$~GeV. For the reduced
signal strength
of the Higgs to di-photon mode, $R_{\gamma\gamma}$, we
use $0.23\leq R_{\gamma\gamma}\leq 1.31$, the latest CMS results at
2$\sigma$~\cite{CMS:yva}\footnote{For ATLAS the same
limit including all systematics is $0.95\leq R_{\gamma\gamma}\leq 2.55$ \cite{ATLAS:2013oma,ATLAS:2013wla}.}.
The remaining reduced signal strengths are also constrained 
within their respective $2\sigma$ ranges according
to the CMS results of Ref.~\cite{CMS:yva} (see Refs.~\cite{ATLAS:2013mma,ATLAS:2013wla} 
for the equivalent ATLAS results).
A bound on the branching ratio for invisible
Higgs decay~\cite{Espinosa:2012vu,ATLAS:2013pma,
CMS:2013bfa,CMS:2013yda,Belanger:2013kya,Ellis:2013lra,Belanger:2013xza
,CMS:1900fga} i.e., BR$(h_{\rm SM} \to \neut_1 \neut_1)$ has
also been considered in our analysis. 
Notice that imposing these measurements indirectly entails a
strong bound on the non-standard decay modes of the SM-like Higgs 
boson~\cite{Belanger:2013kya}, which in our case affects 
BR$(\h2\to \h1\h1)$ and BR$(\h2\to \a1\a1$).
\begin{table}[tb!]
  \begin{center}
    \begin{tabular}{|c|c|}
      \hline
      Parameter & Range
      \\
      \hline
      $\tan\beta$ & $3 - 20$ \\
      $\lambda$& $0.1- 0.7$\\
      $\kappa$& $0.01- 0.6$  \\
      $A_{\lambda}$& $0 - 1000$ \\
      $A_{\kappa}$& $-100 - 100$ \\
      $\mu$& $110 - 300$ \\
      $M_1$& $200 - 500$ \\
      \hline
    \end{tabular}
    \caption{\label{parameters}
    Ranges of variation in the seven parameters used in the scan. 
Masses and trilinear terms are given in GeV units. All the parameters are 
defined at the EW scale.}
  \end{center}
\end{table}

Some low-energy observables also have an important impact in the allowed regions of
the NMSSM. We have implemented the recent measurement of the branching ratio of the $B_s\to \mu^+\mu^-$ process
by the LHCb \cite{Aaij:2013aka} and CMS \cite{Chatrchyan:2013bka} collaboration, which collectively yields
$1.5\times 10^{-9}< {\rm BR}(B_s\to \mu^+\mu^-)< 4.3\times 10^{-9}$ at
95\% CL~\cite{Galanti-talk}. 
For the $b\to s\gamma$ decay, we require the 2$\sigma$ range
$2.89\times 10^{-4}< {\rm BR}(b\to s\gamma)< 4.21\times
10^{-4}$, which takes into account
theoretical and experimental uncertainties added in quadrature \cite{Ciuchini:1998xy,D'Ambrosio:2002ex,Misiak:2006zs,
Misiak:2006ab,Amhis:2012bh}.
We also impose $0.85\times 10^{-4}<$BR$(B^+ \to \tau^+ \nu_\tau)<2.89\times 10^{-4}$ \cite{Lees:2012ju}.
We do not impose any constraint on the SUSY contribution to
the muon anomalous magnetic moment, $a_{\mu}^{\rm SUSY}$. As already
emphasized in Refs.\,\cite{Endo:2011gy,Endo:2013bba}, the regions of
the parameter space that lead to a 125~GeV Higgs generally result in a
small $a_{\mu}^{\rm SUSY}$ which is in tension with  experimental
results using $e^+e^-$ experimental data
\cite{Bennett:2006fi,Hagiwara:2011af}. However, if tau data is used,
the discrepancy is smaller \cite{Davier:2010nc}.

We finally require the lightest neutralino to be the LSP and set
an upper bound on its relic abundance, $\Omega_{\neut_1} h^2<0.13$,
consistent with the latest Planck results~\cite{Ade:2013zuv}.
We also impose limits on the spin-independent neutralino-nucleon
scattering cross section using the most recent experimental results
\cite{Angle:2011th,Aprile:2012nq,Ahmed:2009zw,Ahmed:2011gh,Agnese:2013rvf}.

In our scan, we have built a likelihood function, whose 
parameters are the neutralino relic density and $m_{\h2}$, which are taken 
as gaussian probability distribution functions around the measured values with 
$2\sigma$ deviations. The lightest pseudoscalar mass is also incorporated in the 
likelihood in such a way that masses below  $m_{a^0_1}<20$ GeV are favoured 
(although heavier masses are not excluded). This likelihood function is 
used by \textsc{MultiNest 2.9} to generate 
Markov Chain Monte Carlo (MCMC) chains and find regions of the 
parameter space which maximize the likelihood. Using \textsc{MultiNest} allows 
us to explore the parameter space of the model more efficiently, since relatively 
few evaluations are needed to converge to regions which maximize the likelihood.

\begin{figure}[t!]
  \begin{center}
    \includegraphics[width=0.48\textwidth]{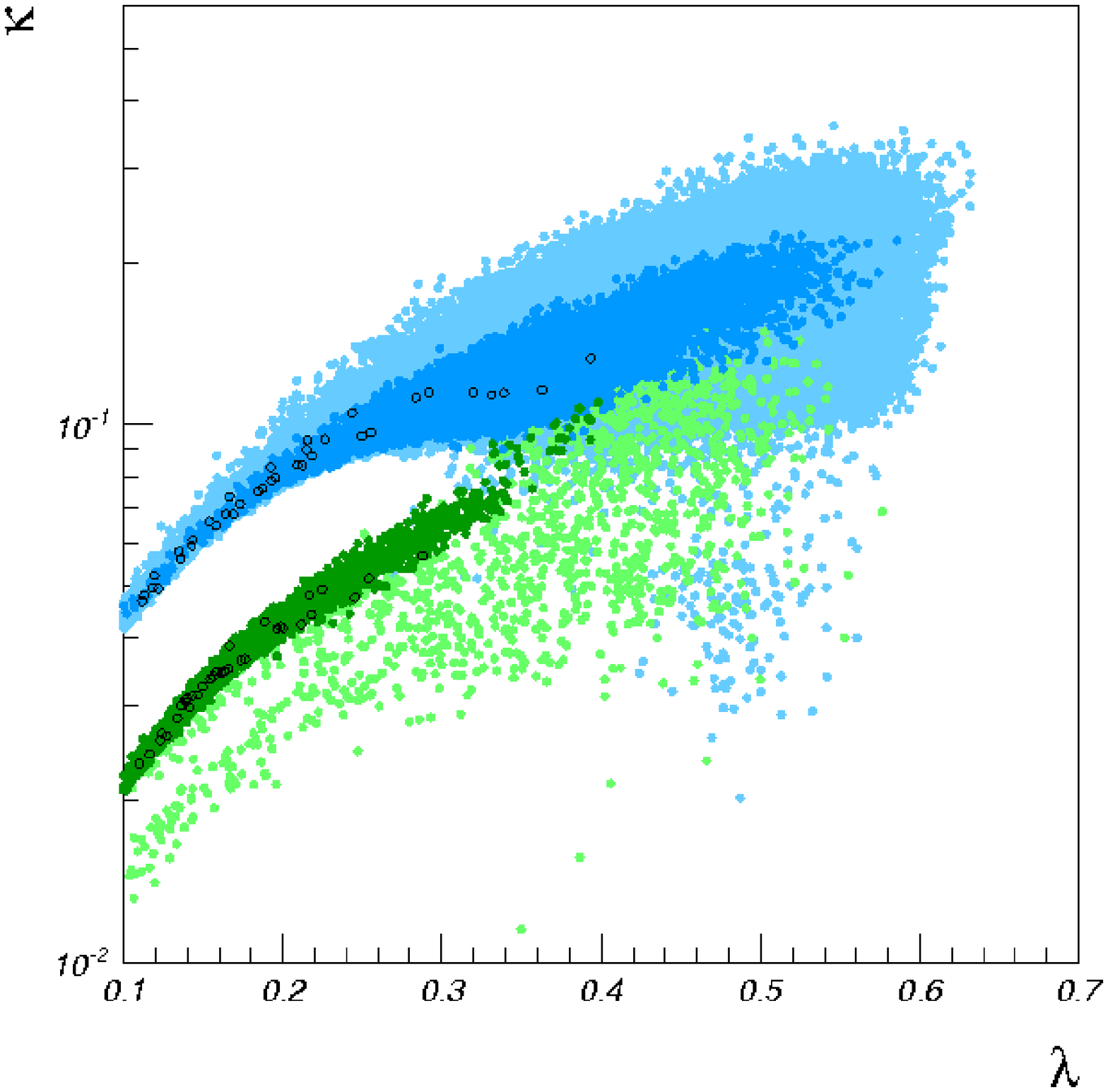}
    \includegraphics[width=0.48\textwidth]{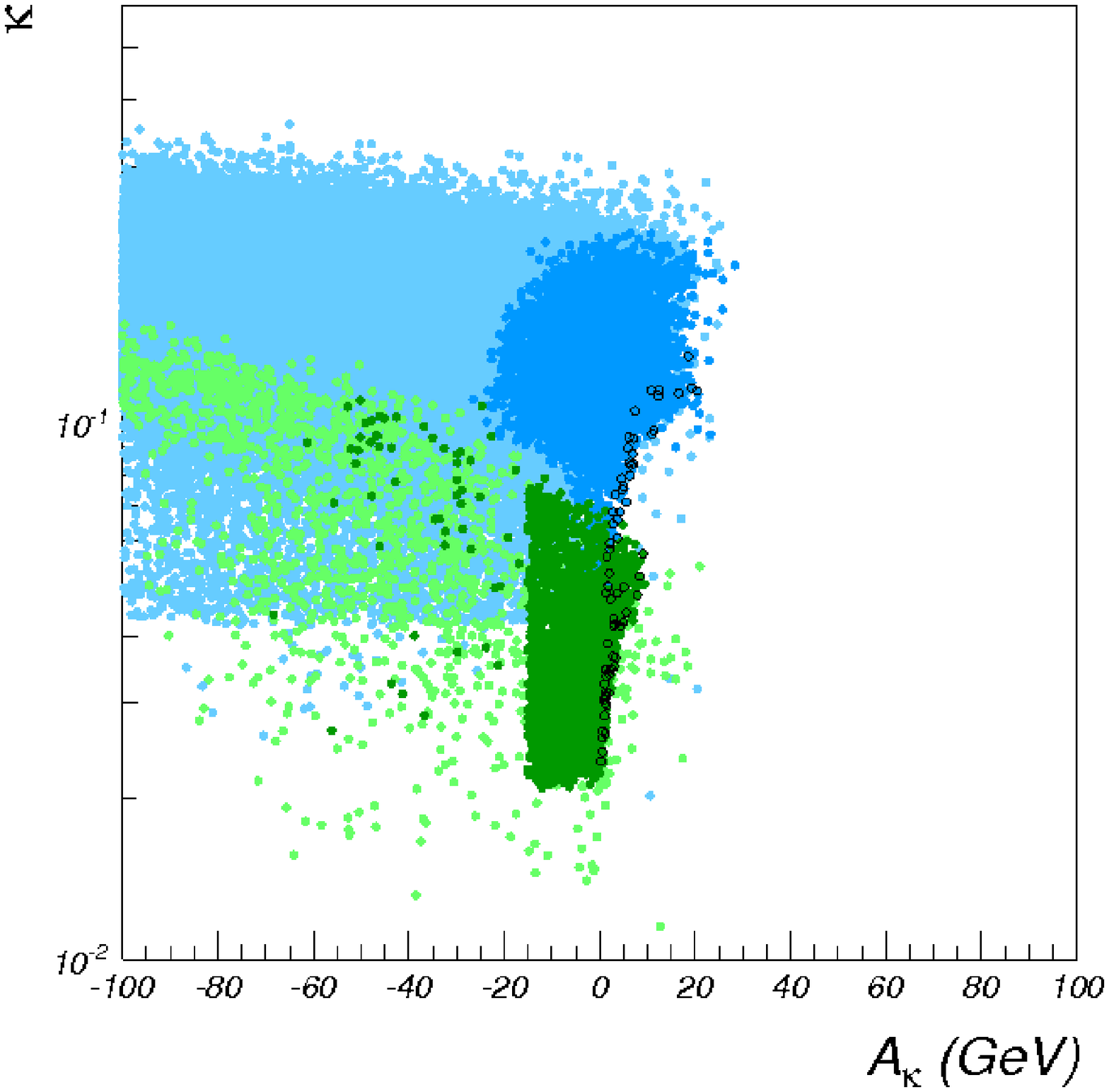}\\
    \includegraphics[width=0.48\textwidth]{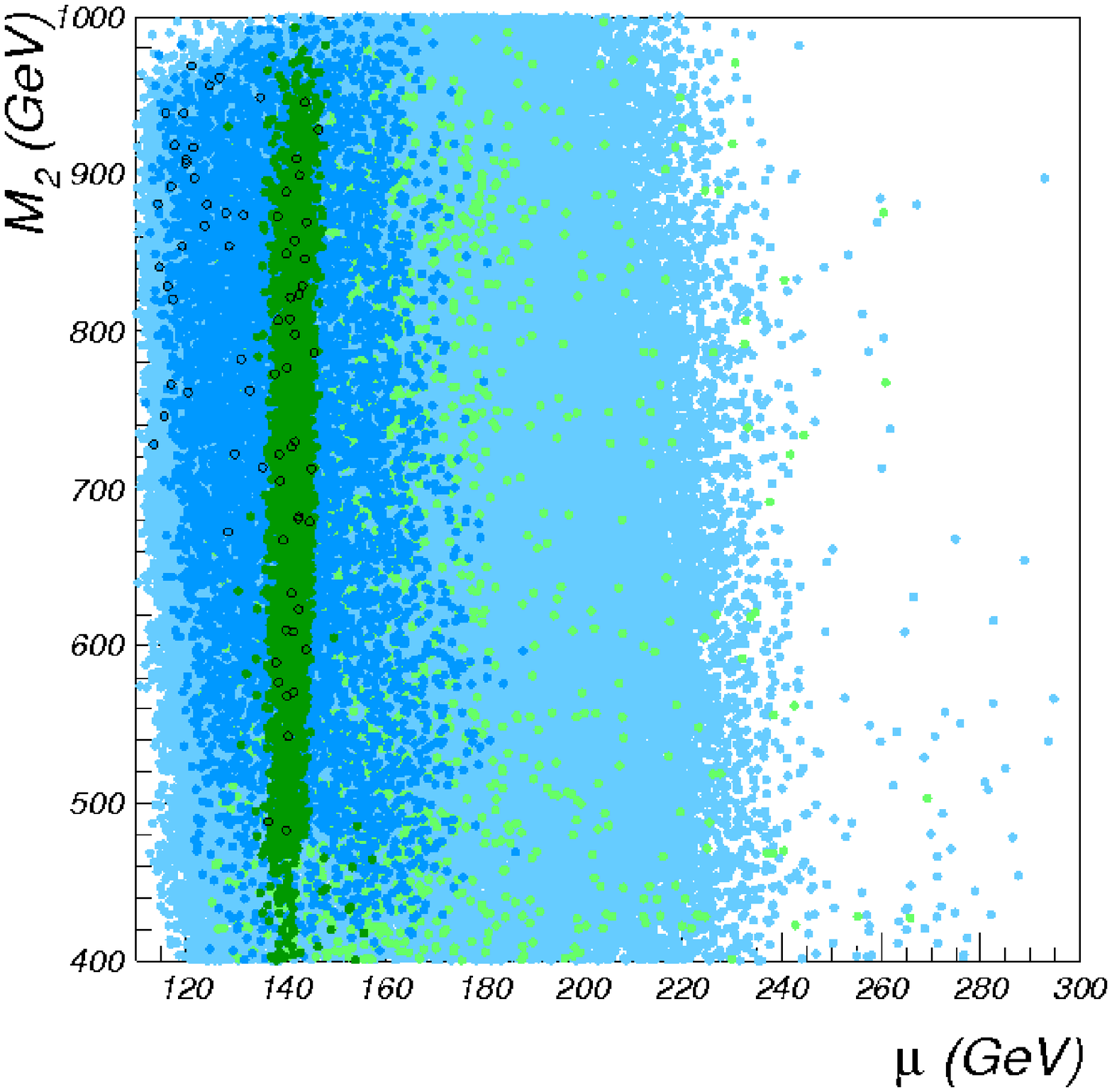}
    \includegraphics[width=0.48\textwidth]{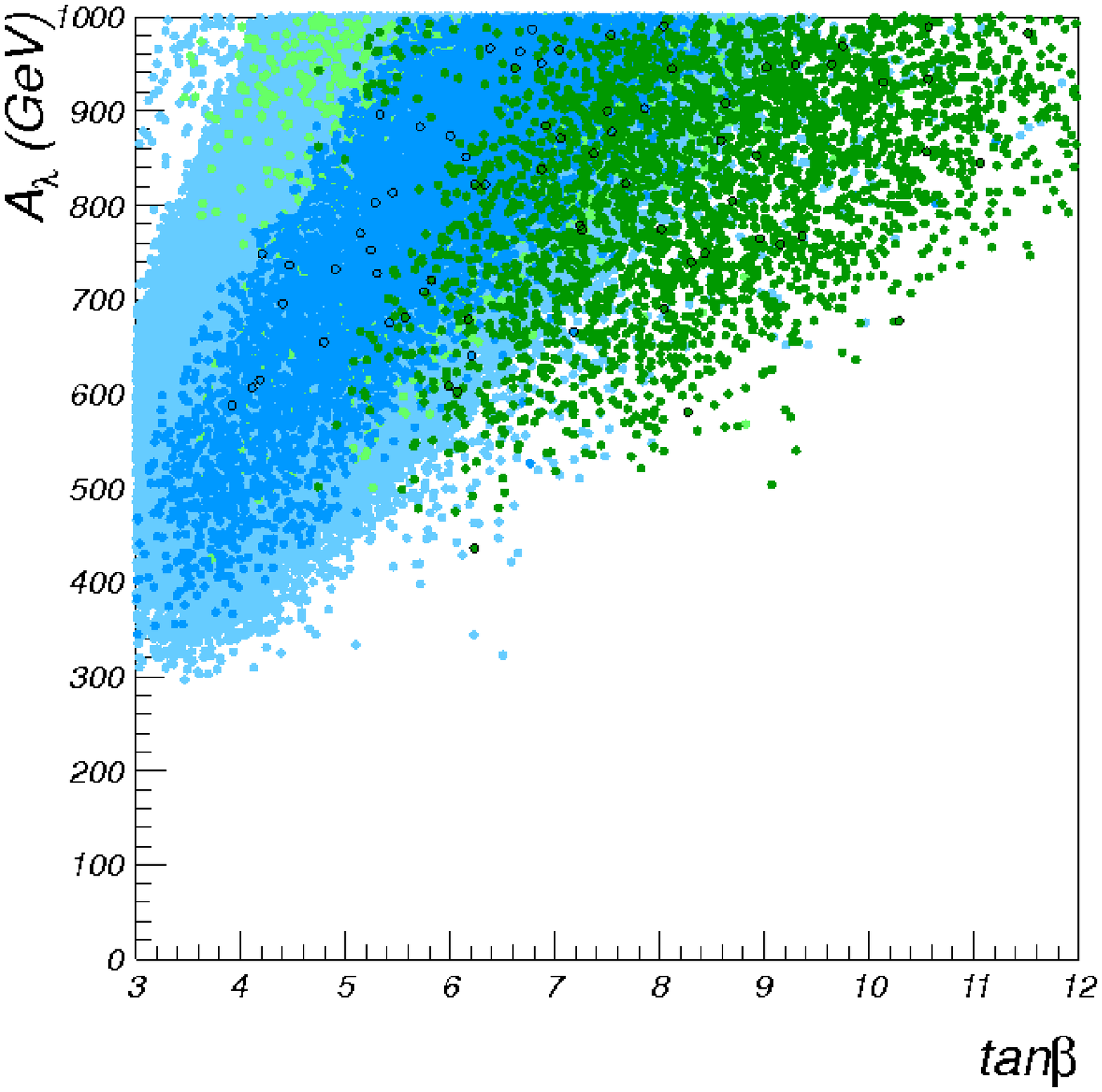}
  \end{center}
  \caption{Scatter plot of the allowed regions in the different parameters used as 
inputs for the scan after all experimental constraints are included. Blue dots 
represent points for which $m_{\h1}>m_{\h2}/2$, whereas green dots are those in 
which $m_{\h1}<m_{\h2}/2$. Darker points correspond to solutions for which 
BR$(\neut_2\to\neut_1 \a1)>0.5$ or BR$(\neut_3\to\neut_1 \a1)>0.5$. On top of this, 
black circles correspond to points with $m_{\a1}<10$~GeV.
}
  \label{fig:scan}
\end{figure}

In Fig.~\ref{fig:scan}, we show the scatter plot corresponding to
different combinations of the input parameters that pass all the experimental constraints.
We distinguish two scenarios according to the masses of the Higgs sector. Blue
dots correspond to points in the parameter space with $m_{\h1}>m_{\h2}/2$, while
the green ones are those with $m_{\h1}<m_{\h2}/2$. Since we are interested in
neutralino/chargino pair production and its final decay through
the $\neut_{2,3}\to\neut_1\a1$ decay channels, we have denoted with
darker colours the points for which BR$(\neut_2\to\neut_1 \a1)>0.5$ 
or BR$(\neut_3\to\neut_1 \a1)>0.5$. On top of this, points in which 
the mass of the lightest pseudoscalar is
lighter than 10 GeV are represented by means of black circles.

In the $(\lambda,\,\kappa)$ plane (top left panel),
points with a lighter $\h1$ (green points) are found in the region of small
values of $\kappa$, as in these areas their singlet component is 
sizable and collider constraints can be avoided.
Similarly, very light pseudoscalars accumulate  towards small values of
$\kappa$, and $A_{\kappa}$, as it can be seen in the top right
panel, where the $(\kappa,\,A_\kappa)$ plane is represented. The
presence of a very light pseudoscalar Higgs in the NMSSM requires a 
tuning of some of the parameters so that either the
U(1)$_R$ or U(1)$_{PQ}$ symmetry of the model is recovered and
then this light pseudoscalar would correspond to the pseudo-Goldstone boson of
the symmetry~\cite{Dobrescu:2000yn,Gunion:2005rw}. In our scan, the
smaller values of the pseudoscalar mass are obtained when
$\kappa,\,A_\kappa\to0$, for which the U(1)$_{PQ}$ symmetry is
quasi-restored \cite{Dobrescu:2000yn}.

\begin{figure}[t!]
  \begin{center}
      \includegraphics[width=0.48\textwidth]{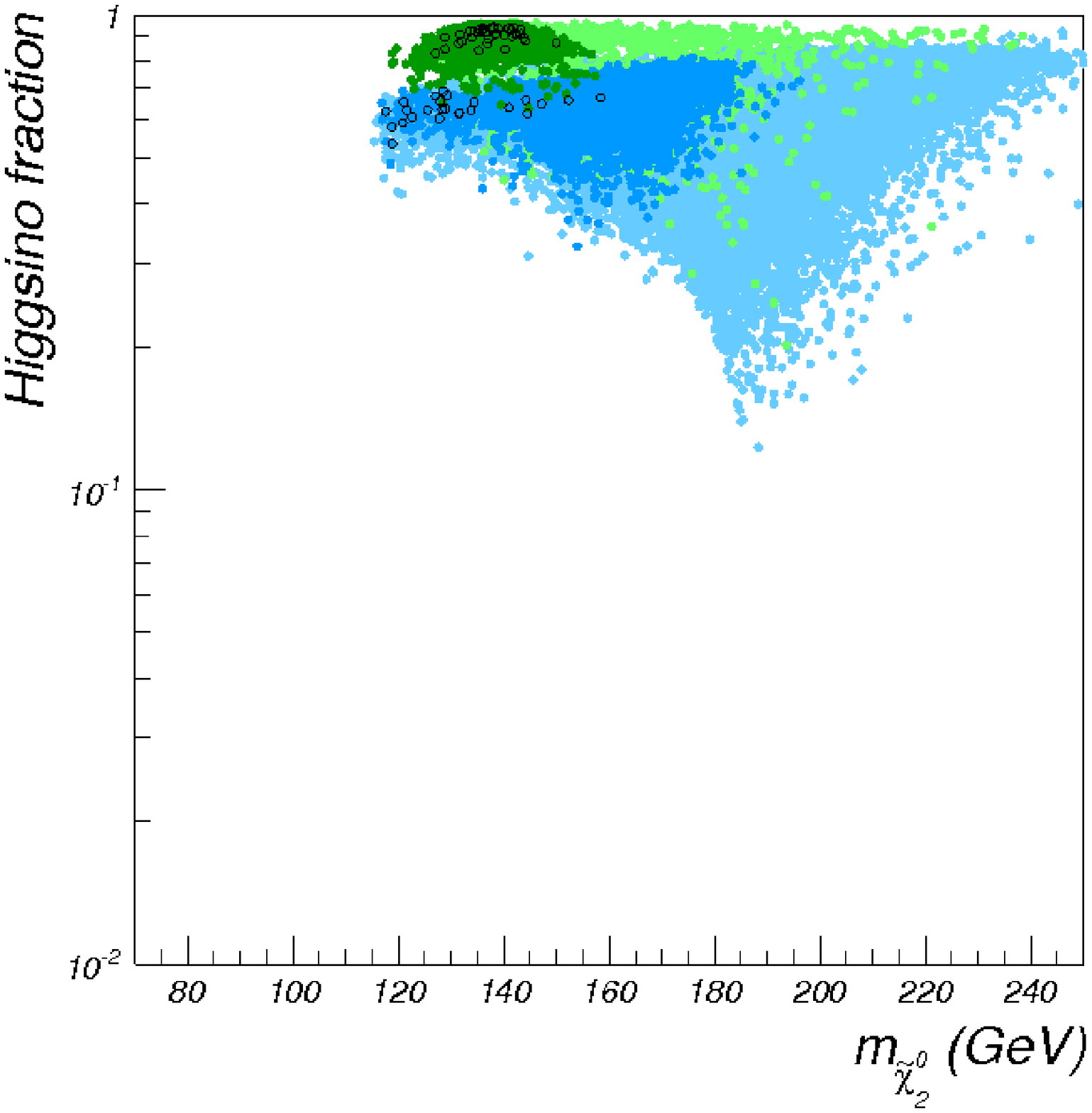}
      \includegraphics[width=0.48\textwidth]{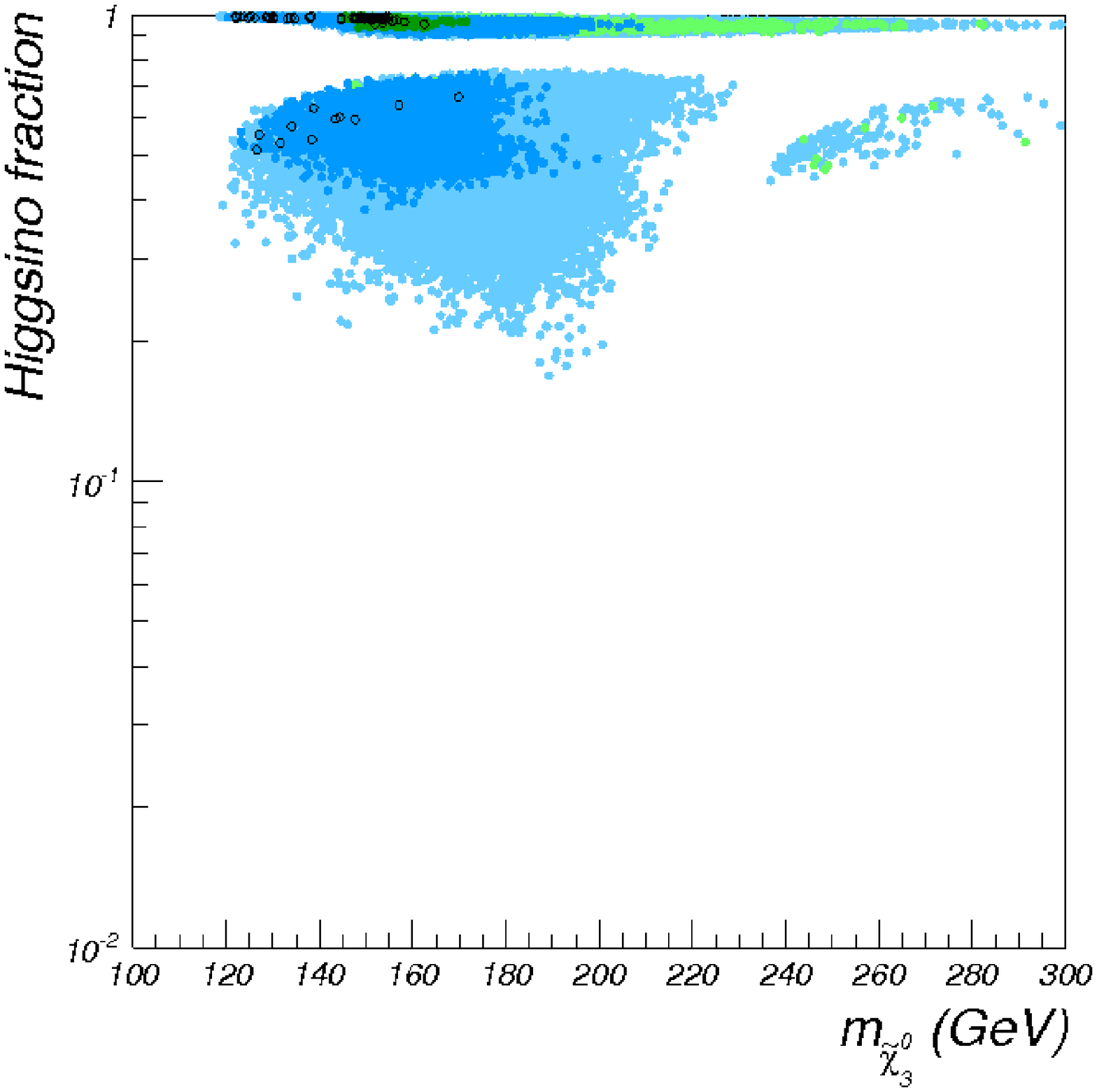}
  \end{center}
  \caption{Higgsino fraction of $\neut_2$ (left) and $\neut_3$ (right) as 
a function of their mass. The colour code is the same as in Fig.\,\ref{fig:scan}.}
\label{fig:higgsino23}
\end{figure}

In the $(\mu,\,M_2)$ plane (bottom left panel), we can observe that in
general the $\mu$ parameter is small. For most of the points, the
hierarchical structure $\mu\lsim M_1<M_2$ is obtained. 
One should note that if gaugino universality at the 
grand unified theory scale is assumed,
all points below $M_2 < 400$ GeV are excluded by the LHC lower
limit on gluino mass. However, we have verified that
if the universality relation between $M_2$ and $M_3$ is broken (setting $M_3$
high so that $m_{\widetilde g} \ge 1.2$ TeV \cite{ATLAS-susy13}) viable points 
for $M_2\leq 400$ GeV can be obtained. This is 
a consequence of the fact that gluino mass usually appears 
in higher order calculations. 

The lightest
neutralino $\neut_1$ is therefore mostly singlino (as a consequence of
the smallness of the $\kappa$ parameter) but with a Higgsino admixture,
which helps to raise its mass and increases for the points
with $m_{\h1}>m_{\h2}/2$.
The points with a sizable BR$(\neut_{2,3}\to\neut_1 \a1)$ occur for
small values of the $\mu$ parameter since this leads to Higgsino-like
$\neut_{2,3}$ and $\widetilde\chi_1^\pm$. Although $M_2$ is generally
large in the scenario with heavier $\h1$, it can be as small as $\sim
300$~GeV in the cases with $m_{\h1}<m_{\h2}/2$. Since this also
implies a small $M_1$, the second and third lightest neutralinos can
also have a non-negligible bino composition.
The Higgsino fractions of $\neut_2$ and $\neut_3$ are plotted as a
function of their mass in Fig.\,\ref{fig:higgsino23}, where we observe
that a large population of points in the parameter space favours
Higgsino-like $\neut_{2,3}$.

Finally, in the $(A_{\lambda},\,\tan\beta)$ plane, we can see
that small values of $\tan\beta$ ($\lesssim 12$) are preferred for the
range used in $A_{\lambda}$. This is useful in order to avoid
constraints on some flavour observables, such as BR($\bmumu$).

\begin{figure}[t!]
  \begin{center}
    \includegraphics[width=0.48\textwidth]{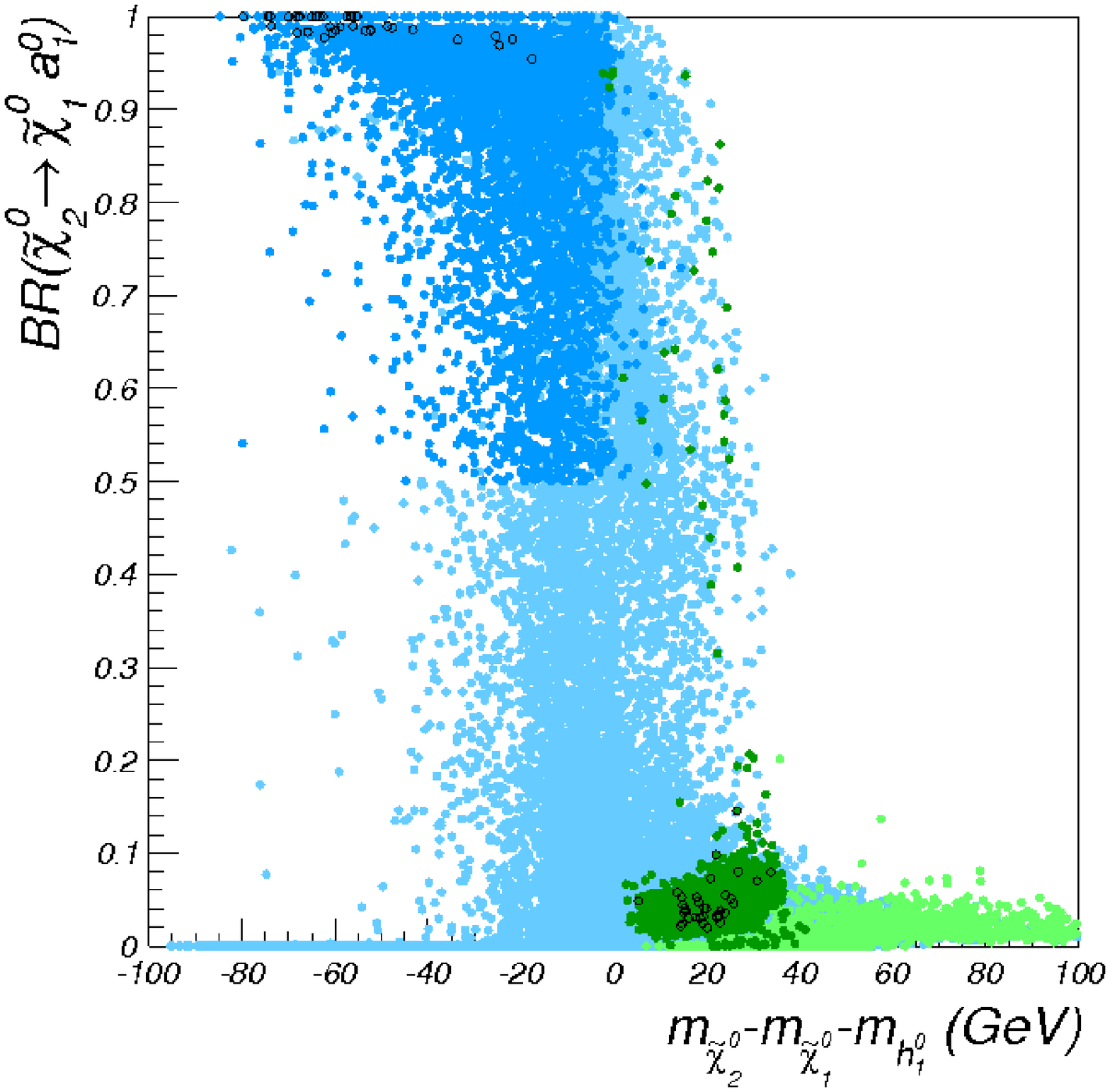}
    \includegraphics[width=0.48\textwidth]{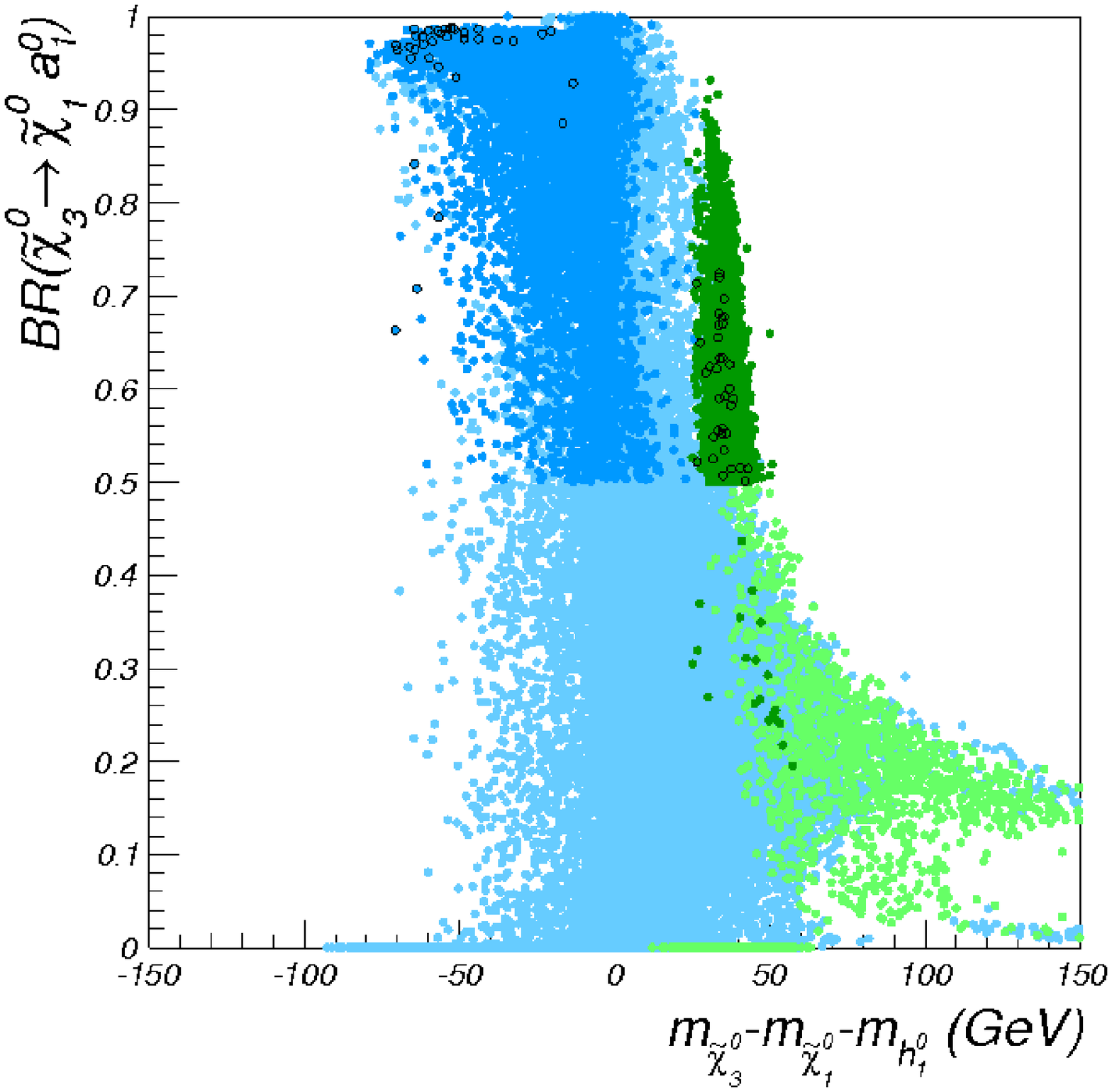}\\
  \end{center}
  \caption{ Left: Br$(\neut_2\to\neut_1 \a1)$ as a function of
    $m_{\neut_2}-(m_{\neut_1}+m_{\h1})$. Right: Br$(\neut_3\to\neut_1
    \a1)$ versus $m_{\neut_3}-(m_{\neut_1}+m_{\h1})$. The colour code
    is the same as in Fig.~\ref{fig:scan}.
}
  \label{fig:br}
\end{figure}

In Fig.\,\ref{fig:br}, we represent the resulting 
BR$(\neut_{2,3}\to\neut_1 \a1)$ as a function of the mass 
difference $m_{\neut_{2,3}}-(m_{\h1}+m_{\neut_1})$.
The alternative decay $\neut_{2,3}\to\neut_1 \h1$ is
kinematically open for $m_{\neut_{2,3}}-(m_{\h1}+m_{\neut_1})>0$. When
this happens, it generally dominates the neutralino decay width and we
obtain small values for BR$(\neut_{2,3}\to\neut_1 \a1)$. This is a
consequence of a relative sign in the corresponding couplings 
for our choice of signs for $\lambda$ and $\kappa$ \cite{Stal:2011cz}.
In the scenario with $m_{\h1}<m_{\h2}/2$, this condition is particularly
constraining since the CP-even Higgs is lighter, and for this reason
we obtain less viable points in this scenario (green points). Still,
we found some solutions featuring BR$(\neut_2\to\neut_1 \a1)>0.5$,
even when the pseudoscalar is very light.
On the contrary, points with heavy $\h1$ (blue points) are more easily
obtained.

\begin{figure}[t!]
  \begin{center}
    \includegraphics[width=0.48\textwidth]{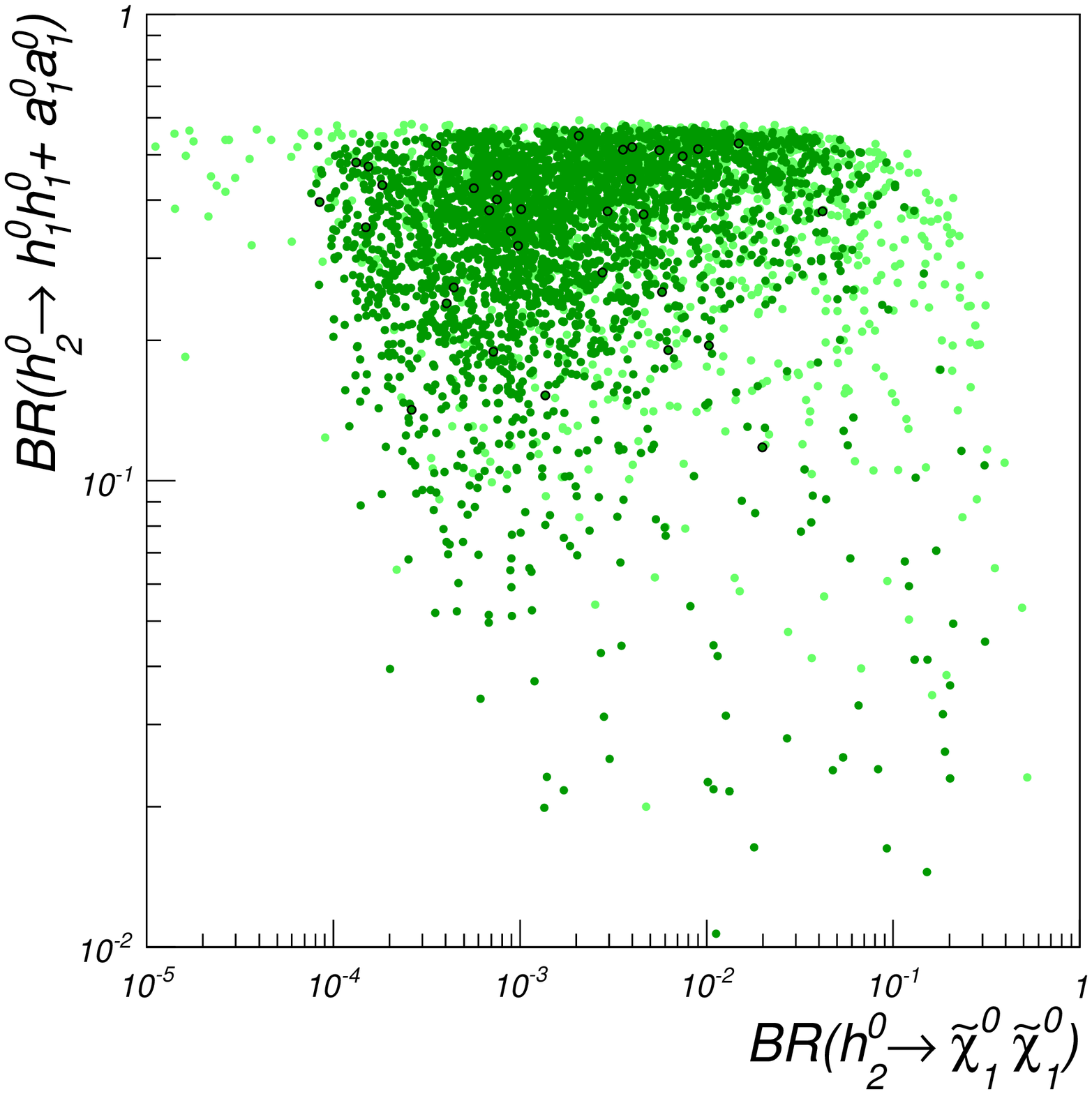}
    \includegraphics[width=0.48\textwidth]{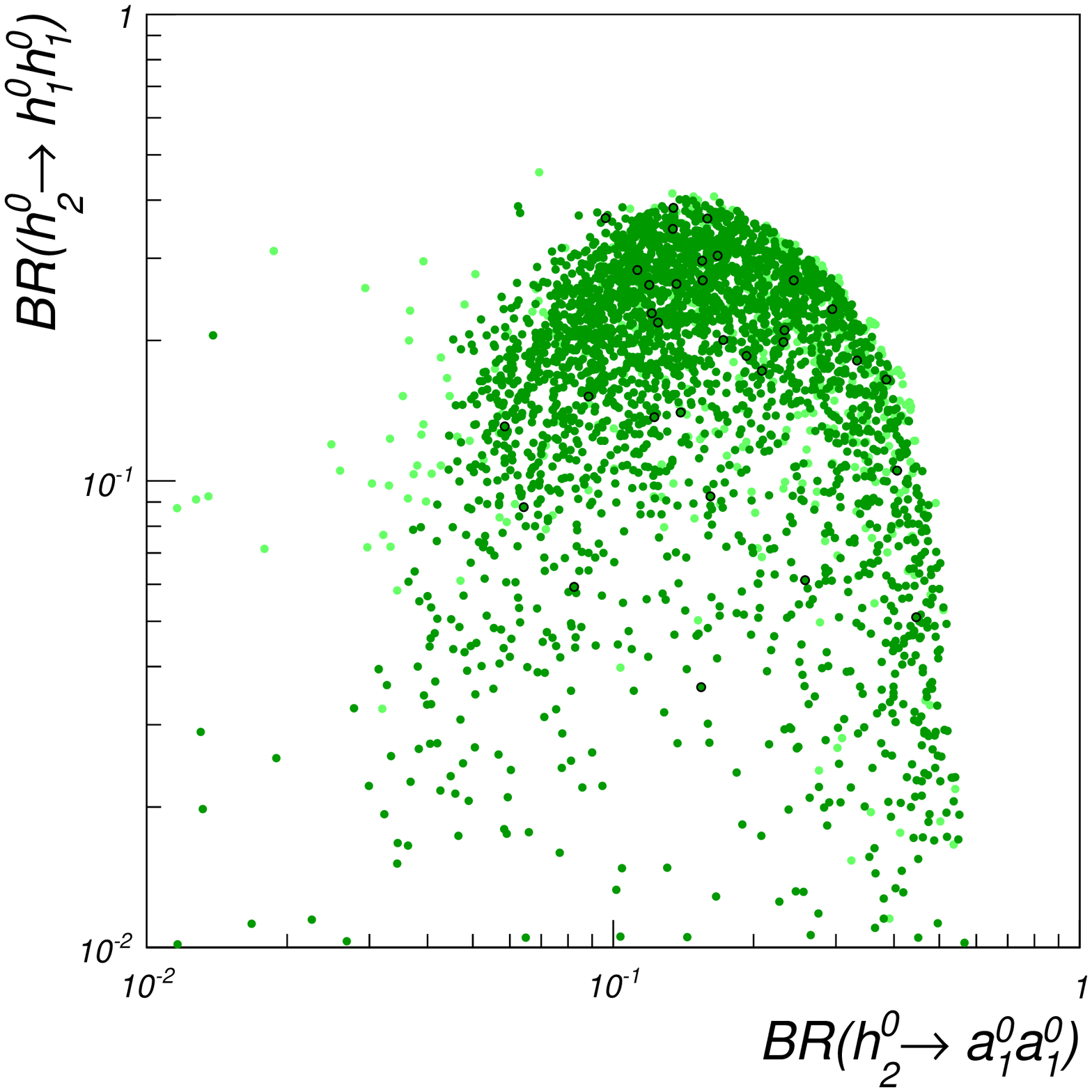}
  \end{center}
  \caption{Left: scatter plot of BR$(\h2\to\h1\h1)+$BR$(\h2\to\a1\a1)$ versus BR$(\h2\to\neut_1\neut_1)$.
Right: scatter plot of BR$(\h2\to\h1\h1)$ versus BR$(\h2\to\a1\a1)$. The colour code
    is the same as in Fig.~\ref{fig:scan} but here we only consider the scenario with $m_{\h1}<m_{\h2}/2$.
}
  \label{fig:h2h2inv}
\end{figure}

As stated earlier, the presence of light $\a1,\,\h1$, and $\neut_1$ can induce 
non-standard Higgs decays. This is particularly important in the regions of the parameter 
space with $m_{\h1} < m_{\h2}/2$ (green points in our plots), since such a light singlet 
$\h1$ is typically associated with a light singlino-like $\neut_1$. Thus, all three decay 
modes $\h2 \to \neut_1\neut_1,
\,\h1 \h1,\, \a1 \a1$ remain kinematically open for this scenario.
In the left plot of Fig.\,\ref{fig:h2h2inv} we have represented the resulting
BR$(\h2\to \h1\h1,\ Ê \a1\a1)$ versus BR($\h2\h2\to\neut_1\neut_1$). 
The two contributions BR$(\h2\to \h1\h1)$ and BR$(\h2\to \a1\a1)$ are plotted separately 
in the right plot of Fig.\,\ref{fig:h2h2inv}. 
The constraints on the reduced signal strengths of the $\h2$ decays imply an indirect 
bound BR$(\h2\to \h1\h1)+$ BR$(\h2\to \a1\a1)+$BR$(\h2\h2\to\neut_1\neut_1)\lesssim0.55$.  
This upper value is relatively high since we are allowing $2\sigma$ deviations in all the 
reduced signal strengths (and in particular on $R_{bb}$).
Notice also that both BR$(\h2\to \h1\h1)$ and BR$(\h2\to \a1\a1)$ can be sizable and 
typically dominate over the invisible decay $\h2\to\neut_1\neut_1$.

\begin{figure}[t!]
  \begin{center}
    \includegraphics[width=0.48\textwidth]{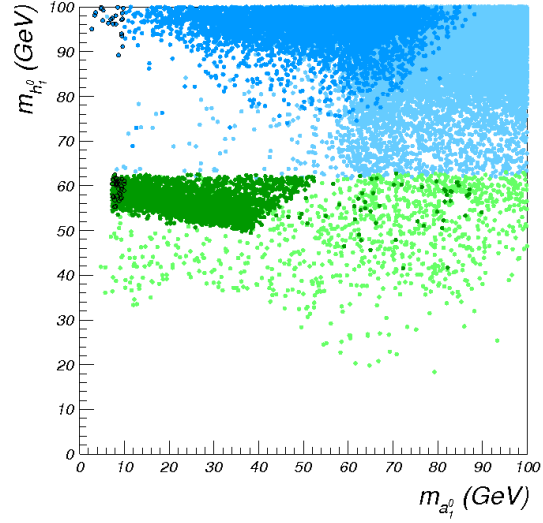}
  \end{center}
  \caption{ Lightest CP-even Higgs mass versus the lightest CP-odd Higgs mass. The color code
    is the same as in Fig.~\ref{fig:scan}.
}
  \label{fig:higgsmass}
\end{figure}

Finally, in Fig.\,\ref{fig:higgsmass},
we represent the values of the lightest CP-even Higgs mass 
versus the lightest CP-odd Higgs mass. Notice that
there is a large population of points in a square region in the upper
right corner. This area satisfies $m_{\h1}>m_{\h2}/2$ and
$m_{\a1}>m_{\h2}/2$, and therefore, $\h1$ and $\a1$ do not alter the
branching ratios of the SM-like Higgs. Outside of this region the constraints on
the properties of a SM-like $\h2$ are very stringent. When we demand a very light
pseudoscalar and a sizable BR$(\neut_{2,3}\to\neut_1\a1)$, two classes
of scenarios are left, those with $m_{\h1}\sim100$~GeV and others with
$m_{\h1}\lesssim 60$~GeV.



In order to proceed with the collider analysis, we have chosen two
points in the parameter space which are representative of the
different patterns of the Higgs spectrum considered in our work. In
particular, the benchmark point BP1 corresponds to an example in which
the lightest CP-even Higgs has a mass $m_{\h1}\sim 98$~GeV, whereas
the other benchmark point BP2 illustrates a case with
$m_{\h1}<m_{\h2}/2$. The input values of the NMSSM parameters defining
these points are given in Table\,\ref{tab:bp}, together with the
corresponding mass spectrum\footnote{Out of convenience, the
value of the top trilinear parameter $A_t$ in these benchmark points
is slightly different from the fixed one used for the 
scan. However, this does not affect our conclusions.}.
These points are similar to those studied in a previous
analysis~\cite{CGP1} in the context of scalar Higgs decays
$h\to\a1\a1$.
It should be noted that in order to obtain such a light $\a1$ 
the input parameters have to be carefully tuned.


 \begin{table}[t!]
 \begin{center}
 {\small \begin{tabular}{|cc| c |c| }
 \hline
 &  & BP1 &  BP2  \\
 \hline
 &$\tan\beta$&5&5 \\
 &$\la,\,\ka$&0.285, 0.114& 0.286, 0.0844\\
 &$A_\la,\,A_\ka$ &660, 13.8& 820, 14.35\\
 &$M_{{\widetilde L}_{i}},\,M_{{\widetilde e^c}_{i}}$
 &300, 300& 300, 300\\
 &$M_{{\widetilde Q}_{i}},\,
 M_{{\widetilde u^c}_{i}},\,
 M_{{\widetilde d^c}_{i}}
 $&1000, 1000, 1000& 1000, 1000, 1000 \\
 &$\mu$&123.0&123.5\\
 &$M_1,\,M_2,\,M_3$&480, 960, 2880& 250, 500, 1500\\
 &$A_{\tau},\,A_{b},\,A_{t}$&$-1600$, 1000, 1850& $-1600$, 1000, 1250\\
  \hline
 &$m_{\h1},\,m_{\h2},\,m_{\h3}$&97.7, 125.5, 662.4& 62.0, 125.6, 739.6 \\
 &$m_{\a1},\,m_{\a2}$&6.3, 660.8& 7.6, 738.4 \\
 &$m_{h^\pm},\,m_{\wt \chi^\pm_1},\,
 m_{\wt \chi^\pm_2}$ &664.3, 122.6, 965.2& 738.8, 118.1, 522.3\\
 &$m_{\wt \chi^0_1},\,m_{\wt \chi^0_2},\,m_{\wt \chi^0_3}$
 &84.4, 136.0, 140.9& 63.8, 125.3, 139.0\\
 &$m_{\wt t_1},\,m_{\wt t_2},\,
 m_{\wt b_1},\,m_{\wt b_2}$ &644.0, 1048.5, 858.7, 861.3
 &950.6, 1134.9, 1037.0, 1038.6 \\
 &$m_{\wt \tau_1},\,m_{\wt \tau_2}$&296.5, 309.5&296.5, 309.5\\
 &$m_{\wt g}$&2768.4&1496.0\\
  \hline
 \end{tabular}}
 \end{center}
 \caption{\label{tab:bp}
 Model parameters that define our choice of benchmark points and
 resulting spectrum.
 The top-quark pole mass is set to 173.5 GeV and
 ${m_b}^{\overline{\rm MS}}(m_b)=4.18$ GeV. All the masses are given
 in GeV.
 }
 \end{table}

\begin{table}[t!]
  \begin{center}
{\small    \begin{tabular}{|c| c |c| }
      \hline&&\\[-2mm]
      & BP1 &   BP2  \\[2mm]
      \hline&&\\[-2mm]
      Br$(\cha_1\to \ell^\pm \nu_\ell \neut_1)$& 0.11 & 0.11 \\[2mm]
      Br$(\neut_2\to\neut_1 \a1)$ &1.00 & 1.00 \\[2mm]
      Br$(\neut_3\to\neut_1 \a1)$, Br$(\neut_3\to\neut_2 \a1)$& 0.98,
      0.0  & 0.76, 0.20 \\[2mm]
      \hline&&\\[-2mm]
      Br$(\a1 \to \tau^+ \tau^-)$
      & 0.93 & 0.92 \\[2mm]
      \hline
    \end{tabular}}
  \end{center}
  \caption{\label{table:br}
    Relevant branching fractions for multi-lepton
    search channels in the two benchmark points. 
    }
    \end{table}

In both benchmark points (BP1, BP2) the lightest neutralino, $\neut_1$ is mainly singlet-like
$(56.2\%,\,66.9\%)$ with a sizable Higgsino composition $(43.3\%,\,30.5\%)$.
Regarding $\neut_{2,3}$ and $\cha_1$, they are mostly Higgsino and
light (due to the smallness of the $\mu$ parameter and the relatively
large values of $M_1$ and $M_2$), and thus their productions are
greatly enhanced.

The pair production of Higgsino-like $\neut_2$, $\neut_3$, and
$\cha_1$ can lead to multi-lepton final states through decay chains in
which very light pseudoscalars are produced. More specifically, in the
scenarios under study, the second and third lightest neutralinos decay
as, $\neut_{2,\,3}\to\neut_1 + \a1$,
$\neut_{3}\to\neut_2 + \a1\to\neut_1 + 2 \a1$, and for the range of masses 
considered ($2m_\tau<m_{\a1}<2m_b$), the lightest pseudoscalar predominantly 
decays into a pair of taus, $\a1 \to \tau^+ \tau^-$. 
Notice that this differs from conventional analysis, in which 
sleptons or $Z$ intermediate states are involved in lepton production.
On the other hand, the lighter chargino, $\wt \chi^\pm_1$, mainly decays
through slepton or sneutrino mediated standard modes into $\neut_1 \,
\ell^\pm \, \nu_\ell$.
The corresponding branching ratios for these processes can be found in 
Table\,\ref{table:br}, where we can observe that they are sizable in both benchmark points.

The signals of interest are therefore
\begin{eqnarray}
\label{signal}
\neut_{2,\,3}\cha_1 &\to& \ell^+\ell^-\ell^\pm+\MET, \nonumber \\
\neut_{3}\cha_1 &\to& 2\ell^+2\ell^-\ell^\pm+\MET, \nonumber \\
\neut_{3}\neut_{1} &\to& 2\ell^+2\ell^-+\MET,\nonumber\\
\neut_{2,\,3}\neut_{2,\,3} &\to& 2\ell^+2\ell^-+\MET,\nonumber\\
\neut_{2}\neut_{3} &\to& 3\ell^+3\ell^-+\MET,\nonumber\\
\neut_{3}\neut_{3} &\to& 4\ell^+4\ell^-+\MET.
\end{eqnarray}
The missing energy is associated to the lightest neutralino $\neut_1$
and the neutrinos from $\tau$ or $\cha_1$ decays. 
Since BR$(\a1\to\tau^+\tau^-)\sim 1$,
the fraction of lepton flavors in Eq.\,(\ref{signal}) is 
related to that of tau decays.\footnote{In principle the longer 
decay mode like $\neut_{2,3}\to\neut_1\h1\to\neut_1\a1\a1$ can also 
give rise to interesting multi-lepton final states. In this work, however, 
we only considered the simplest mode of Eq.\,(\ref{signal}) since 
it already gives good statistical significance.}

Notice that in principle one can also consider 
neutralino-chargino production from stop decay.
In fact, as it was argued in Ref.\,\cite{Cheung:2008rh}, this could be
an important production channel if gluinos or stops were light, and
$\wt t_1 \to \neut_2 t,\,\neut_3 t,\, \cha_1 b$ can be enhanced if the
Higgsino components of $\neut_2,\,\neut_3,\,\cha_1$ are large. These
decay chains can
give rise to multilepton signals accompanied by hadronic jets and
missing energy $\wt t_1 \wt t^\ast_1 \to n \ell^+  + n^\prime \ell^- +
n^{\prime\prime}~{\rm jets} + \MET$.
This signal can be important 
when LHC starts 
operating at a higher center of mass energy.
However, given the current lower mass bounds on coloured particles from
the current 8 TeV LHC results, squark or gluino decays are not generally
the main production channels for neutralinos in many points of the parameter
space. This statement is generically true even when squark
masses are around $1$~TeV: although squark pair production can be 
significant, the cascade decays are generally suppressed by the corresponding 
branching fractions on each step.
Thus, contrary to the analysis of Ref.\,\cite{Cheung:2008rh},
we will only consider neutralino/chargino pair production.

\section{Direct production of neutralinos decaying into a light pseudoscalar}
\label{sec:prod}


\begin{table}[t!]
  \begin{center}
{\small       \begin{tabular}{|r|c|c|}
      \hline&&\\[-2mm]
      & BP1 & BP2
      \\[2mm]
      \hline&&\\[-2mm]
      $\sigma_{\widetilde\chi_3^0 \widetilde\chi_{1}^0}+\sigma_{\widetilde\chi_3^0 
\widetilde\chi_{2}^0}+\sigma_{\widetilde\chi_3^0 \widetilde\chi_{3}^0}$
      &
      185.8 & 491.9
      \\[2mm]
      $\sigma_{\widetilde\chi_3^0 \widetilde\chi_1^\pm}$ &
      437.2 & 729.1
      \\[2mm]
      $\sigma_{\widetilde\chi_2^0 \widetilde\chi_{1}^0}+\sigma_{\widetilde\chi_2^0 
\widetilde\chi_{2}^0}$%
      &
      309.8 & ~~~4.3
      \\[2mm]
      $\sigma_{\widetilde\chi_2^0 \widetilde\chi_1^\pm}$ &
      727.2 & 648.7
      \\[2mm]
      \hline
    \end{tabular}}
  \end{center}
  \caption{
    Cross sections in fb calculated with \textsc{Herwig++} for the
    direct production of neutralino and chargino pairs at the 8 TeV
    LHC for our choice of benchmark points.}
  \label{table:production}
\end{table}


In this section, we examine how collider signatures of
Eq.~(\ref{signal}) can be detected by using dedicated object
reconstruction schemes and kinematic variables.
The sparticle mass spectrum and decay widths for the selected
benchmark points are calculated with \textsc{nmssmtools}, whose output
is processed with \textsc{Herwig++ 2.6.3}~\cite{Bahr:2008pv,Arnold:2012fq},
interfaced with CTEQ6L1 parton distribution
functions~\cite{Pumplin:2002vw}, in order to calculate the production
cross-sections.
For the SUSY signals, all possible pair productions of the light
neutralinos $\wt\chi_{1,2,3}^0$ and the charginos
$\wt\chi_1^\pm$ have been considered. The production
cross sections for each benchmark point are given in
Table\,\ref{table:production}, where we can observe that the main
products are neutralino-chargino pairs, i.e., $\wt
\chi_{2,3}^0 \wt\chi_1^\pm$, whereas neutralino pairs
$\wt \chi_{2,3}^0 \wt\chi_{2,3}^0$ are sub-leading.
The differences of the cross section values are mainly due to the
singlet compositions of the light neutralinos. For instance,
the singlino component of $\wt\chi_2^0$ in BP2 is $\sim 30$\%, while it
is only $\sim 3\%$ in BP1, which results in a reduced value of the cross section
for neutralino-pair production, $\sigma_{\wt\chi_2^0
  \wt\chi_{1,2}^0}$, in BP2.

In order to study the feasibility of observing the proposed signal at
the current configuration of the LHC, we have generated Monte Carlo
(MC) event samples of the NMSSM signal of direct neutralino pairs, as
well as neutralino-chargino pairs, for a proton-proton collision at
the center-of-mass energy of 8 TeV using \textsc{Herwig++}.
After performing the parton showering and the hadronization with
\textsc{Herwig++}, the generator-level MC events have been processed
with \textsc{Delphes 3.0.7}~\cite{deFavereau:2013fsa} using a modified CMS
card to obtain the detector-level data. Jets are formed using the
anti-$k_t$ jet clustering algorithm~\cite{Cacciari:2008gp} with the
distance parameter of 0.5.
Then, they are required to have a transverse momentum $p_{\rm T} > 20$ GeV
and a pseudo-rapidity $|\eta| < 2.5$ in the analysis.
The $b$-tagging efficiency is set to be 70\% for a jet with $p_{\rm T}
> 30$ GeV, while the mis-tagging rates are assumed to be 10\% and 1\%
for the $c$-jets and the light-flavor jets, respectively.

For choosing isolated lepton candidates (throughout the text, isolated lepton 
includes $e$, $\mu$, and hadronically-decaying $\tau$, $\tau_h$), the threshold $p_{\rm T}$ of
5 GeV is adopted and the scalar sum of the transverse momenta of whole
charged tracks with $p_{\rm T} > 0.5$ GeV lying in a cone of $\Delta R
= 0.1$ around the candidate has been calculated 
(we will later justify the need for such a small $\Delta R$).
The candidate lepton is accepted as being isolated if the fraction of
the sum to the candidate's transverse momentum is less than 10\%.
This isolated lepton criterion is different from the one used in the
ATLAS~\cite{ATLAS_chi} and CMS~\cite{CMS_chi} analyses to look for the
direct neutralino-chargino pair production.
For instance, the CMS analysis selects the cone size of 0.3 and the transverse momentum 
fraction less than 15\%, in order to pick up both isolated electrons and muons.
A better isolation of the lepton with larger cone size 
appears at the cost of reducing the number of isolated leptons.
However, we find that the criteria adopted by the experimental collaborations often fail to 
capture the signal leptons since the leptons in the signal of interest are likely to be very 
close to the other lepton sharing the same parent pseudoscalar and possess
relatively soft transverse momentum. When adopting the conventional
criterion, we found that the signal is almost hidden in the
backgrounds, and for this reason
we relax the criterion of the cone size while being more strict on the
fraction value. One can further attempt to tune the parameters for the
signals of each benchmarks, however it
is beyond the scope of our work since a concrete knowledge of the detector
performance is necessary for that. Instead of tuning,
the fake leptons originated from the jets are removed by
imposing separate cuts.

\begin{figure}[t!]
  \begin{center}
    \includegraphics[width=0.48\textwidth]{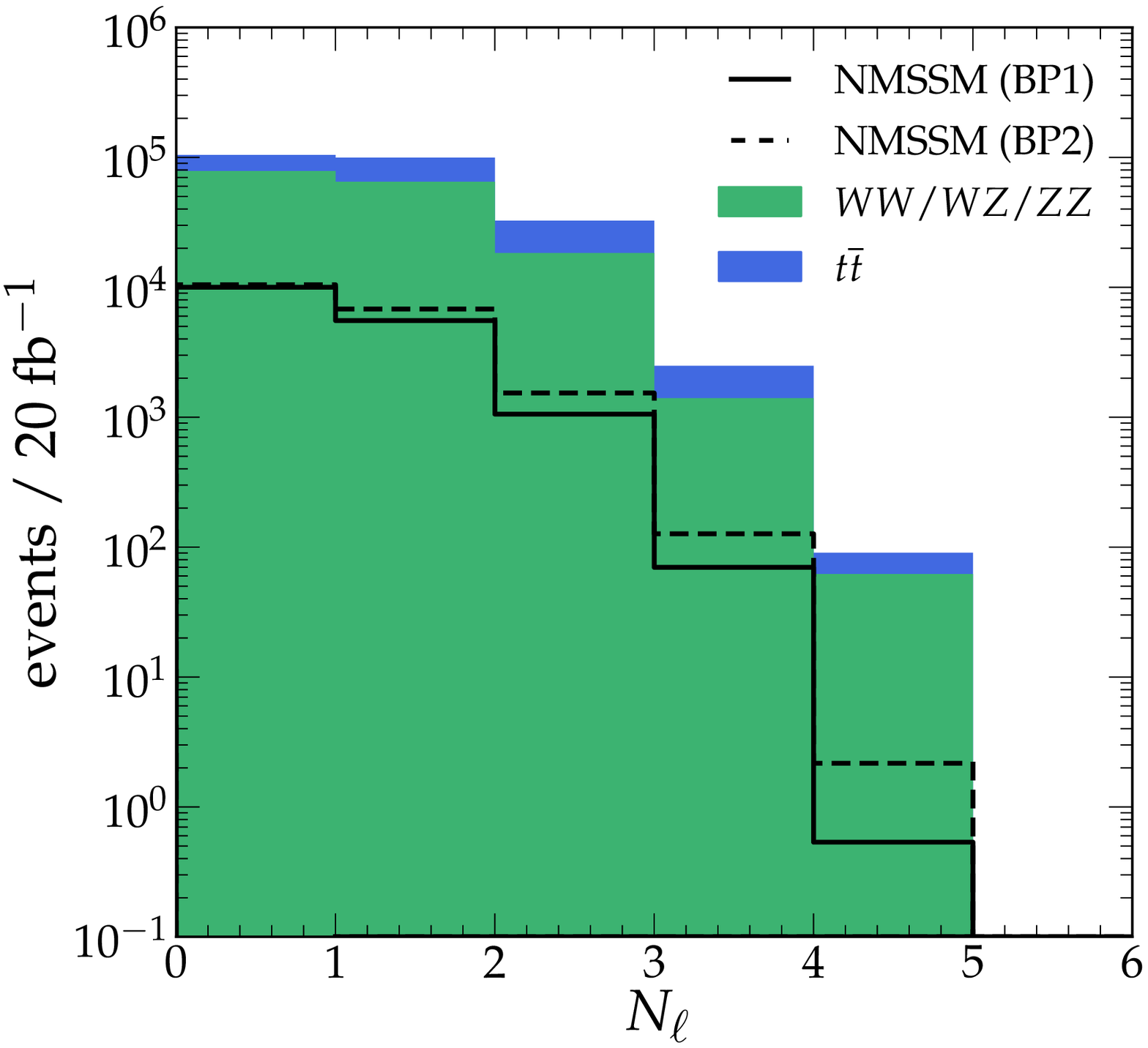}
    \includegraphics[width=0.48\textwidth]{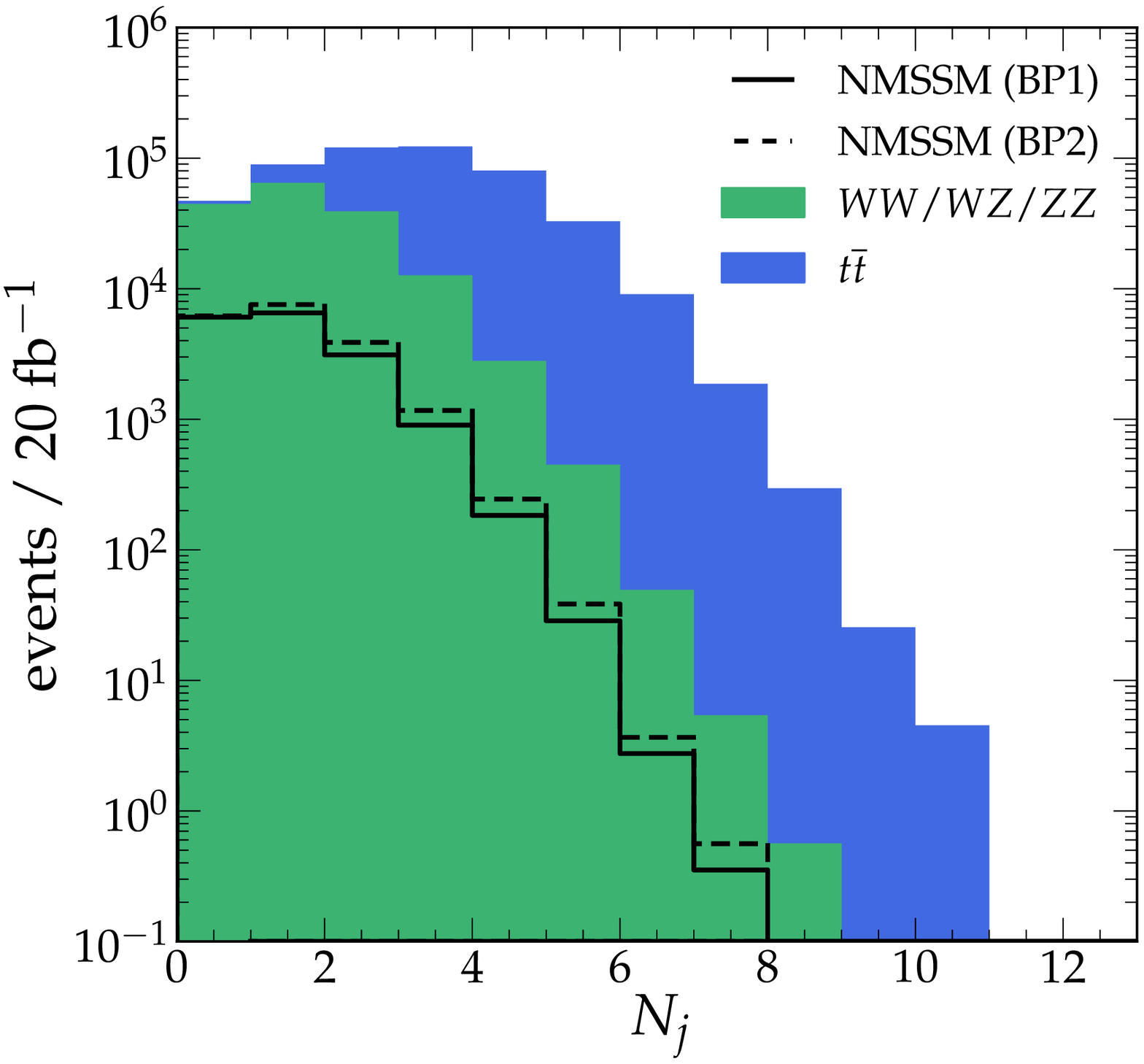}
  \end{center}
  \caption{Number of (left panel) charged leptons satisfying $p_{\rm
      T} > 10$ GeV and $|\eta| < 2.4$ after choosing the 
isolated leptons and (right panel) jets with $p_{\rm T} > 20$
    GeV and $|\eta| < 2.5$.}
  \label{fig:n_lepton}
\end{figure}

The isolated lepton chosen by the criterion is
discarded if its angular separation to the adjacent jet with
$p_{\rm T} > 20$ GeV is within a range of $\Delta R <
0.4$. Furthermore, the event containing an opposite-sign same-flavor
(OSSF) lepton pair with an invariant mass below 12~GeV is rejected to
suppress the low-mass continuum backgrounds. This cut also rejects some
signal events due to the low-mass pseudoscalar, however, the signals
can still pass the cut since they can have different-flavor lepton pairs.
Then, the event is selected for the analysis when it has at least one
isolated electron or muon with $p_{\rm T}^e > 12$ GeV or $p_{\rm T}^\mu > 8$ GeV and $|\eta| < 2.4$.
Since the efficiency for the tau-jet identification is poor and our
analysis largely relies on leptonic tau decays, we use a
conventional criterion for reconstructing the tau-jets with a cone
size of 0.5 and the minimum $p_{\rm T}$ value of 10 GeV. In the
analysis, we select only the tau-jet with $p_{\rm T} > 15$ GeV. In
Fig.~\ref{fig:n_lepton}, we show the lepton and jet multiplicity
distributions for both signals and backgrounds.
The missing transverse momentum $\mathbf{\MPT}$ is defined as the
negative vector sum of the transverse momenta of all the calibrated
calorimetric energy clusters and muon candidates. In SUSY signals,
the main source of the missing energy is the neutrinos from the decays of
taus or charginos, as well as the undetectable neutralino LSP.

The main SM backgrounds consist of the EW diboson ($WW$, $WZ$, $ZZ$) and
triboson ($WWW$, $WWZ$, $ZZZ$), resulting in leptonic final states,
the dileptonic top-pair, $t\bar{t}W/Z$, Drell-Yan (DY), and $Z+{\rm jet}$ processes.
All the background processes except triboson and $t\bar{t}W/Z$, for
which \textsc{MadGraph 5}~\cite{Alwall:2011uj} has been used, are
generated by \textsc{Herwig++}, interfaced with $\textsc{Delphes}$ to
simulate the detector effects and reconstruct the final-state objects.
We use the measured cross section values in the recent CMS analysis
results for the most important SM background processes of
diboson~\cite{CMS:diboson,Chatrchyan:2013oev} and dileptonic
$t\bar{t}$~\cite{CMS:ttbar}, while the values calculated with
\textsc{Herwig++} for the DY and $Z+{\rm jet}$ and \textsc{MadGraph}
for the triboson and $t\bar{t}W/Z$ processes are used when estimating
the backgrounds.

Concerning SUSY backgrounds with conserved $R$-parity,
the dilepton invariant mass around $m_{\a1}$ can be a useful separator.
On the other hand, since the singlino like $\neut_1$ can
be very light, large missing transverse energy might no longer be a good
discriminator to $R$-parity violating models. 
Especially, non-minimal SUSY models with broken $R$-parity,
e.g., the $\mu\nu$SSM \cite{LopezFogliani:2005yw,Escudero:2008jg} 
can accommodate similar light scalars, pseudoscalars
and neutralinos \cite{Fidalgo:2011ky,Ghosh:2012pq} and hence produces
similar final states. 
However, measurable
displaced vertices for the latter class of models could be useful to 
distinguish among these constructions.
Finally, backgrounds arising either from squark/gluino mediated
cascades or from decays of heavier neutralino-chargino pair can be
isolated from the studied signal in terms of final state lepton and jet multiplicity.%

In order to increase the ratio of the signal to the
backgrounds, the following basic event selection cuts are applied.
\begin{itemize}
  \item At least three isolated leptons, $\ell = e,\,\mu,\,\tau_h$, where
    $\tau_h$ denotes the $\tau$-jet, and at least
    one of them is required to have $p_{\rm T} > 20$ GeV.
  \item No $b$-tagged jet.
  \item For electrons and muons, the invariant mass of the
    OSSF leptons
    $m_{\ell^+\ell^-}^{\rm OSSF}$ must satisfy
    $|m_{\ell^+\ell^-}^{\rm OSSF} - m_Z| > 15$ GeV to exclude the
    backgrounds associated with the leptonically-decaying $Z$ boson.
\end{itemize}

\begin{figure}[t!]
  \begin{center}
    \includegraphics[width=0.5\textwidth]{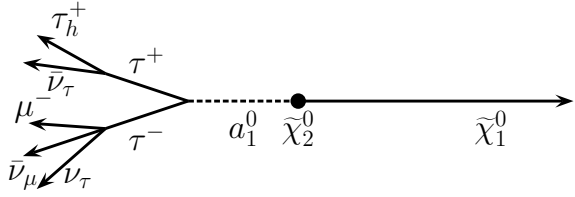}
  \end{center}
  \caption{A kinematic configuration of the signal decay events in the
    rest frame of $\wt \chi_2^0$.}
  \label{fig:kinematic_conf}
\end{figure}

\begin{figure}[t!]
  \begin{center}
    \includegraphics[width=0.48\textwidth]{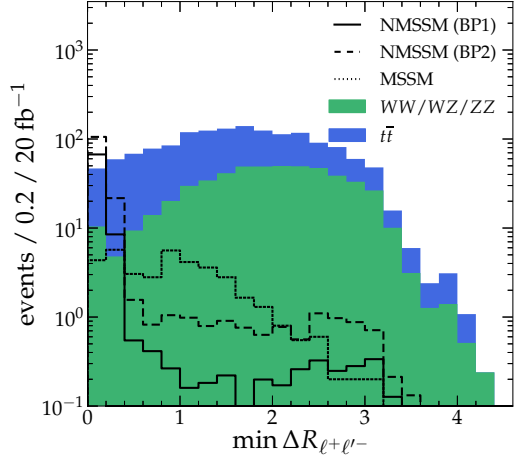}
  \end{center}
  \caption{Distributions of the smallest $\Delta R_{\ell^+\ell^{\prime
        -}}$
    The basic event selection cuts are applied for both signals and
    backgrounds.}
  \label{fig:delta_r}
\end{figure}

In the latest ATLAS and CMS studies, the
neutralinos and charginos are assumed to be practically EW gauginos
and the decays are mediated by on/off-shell sleptons or EW gauge
bosons like
\begin{align}
  \wt \chi_2^0 + \wt \chi_1^\pm
  \to \ell^\pm \wt \ell^{\mp (\ast)} \,\left( Z^{(\ast)} \,
  \wt \chi_1^0 \right)+ \ell^{\prime \pm} \nu_{\ell^\prime} \wt
  \chi_1^0 \to \ell^+ \ell^- \ell^{\prime \pm} + \MET .
\end{align}
In order to interpret the search results, simplified SUSY model points
are considered and the missing energy $\MET$, the dilepton invariant
mass $m_{\ell\ell}$, and the transverse mass $M_{\rm T} \equiv
\sqrt{(|\mathbf{p}_{\rm T}^\ell| + \MET)^2 - |\mathbf{p}_{\rm T}^\ell
  + \mathbf{\MPT}|^2}$ are employed as the main kinematic variables.
In our benchmark scenarios, the final state can be similar to that
in the simplified models,
however, the search strategy should be basically different from those
studies not only because of the fact that the Higgsino-like
neutralinos $\wt \chi_3^0$ as well as $\wt \chi_2^0$
come into play in the SUSY signal productions, but also because of the existence
of the light pseudoscalar that decays predominantly into a pair of tau
leptons.
In particular, due to the light pseudoscalar, the taus are nearly
collinear and the visible final state particles become
relatively soft since a portion of tau energy is carried away by the
neutrinos. In Fig.~\ref{fig:kinematic_conf}, one can see an example of
the kinematic configuration in the rest frame of $\wt \chi_2^0$.
The kinematic configuration often results in the failure of
reconstructing the $\tau$-jet, while the
electron or muon can still have chances to be identified as the isolated
lepton. If at least two isolated leptons ($e,\,\mu$ or $\tau$-jet), sharing the
same parent pseudoscalar $a_1^0$, are successfully identified, their
angular separation $\Delta R_{\ell^+\ell^{\prime -}} \equiv
\sqrt{(\Delta \phi_{\ell^+\ell^{\prime -}})^2 +
  (\Delta \eta_{\ell^+\ell^{\prime -}})^2}$ will turn out to
be small and can be estimated as
\begin{align}
  \Delta R_{\ell^+\ell^{\prime -}} \sim \frac{4m_{\wt \chi_i^0}
    m_{a_1^0}}{m_{\wt \chi_i^0}^2 - m_{\wt \chi_j^0}^2}\ ,
\end{align}
in the case of $\wt \chi_i^0 \to \wt \chi_j^0 a_1^0
\to \wt \chi_j^0 \tau^+ \tau^-$.
Since our basic event selection cuts demand that
there are at least three leptons in the event, all possible
combinations of the opposite-sign leptons are considered, and then
the smallest value of $\Delta R_{\ell^+\ell^{\prime -}}$ is chosen.
Both signal and dominant background distributions are shown in
Fig.~\ref{fig:delta_r}.

A similar final state can arise in a simplified MSSM with the
sparticle mass hierarchy of $m_{\wt \chi_1^0} < m_{\wt
  \tau} < m_{\wt \chi_2^0} < m_{\wt e, \, \wt
  \mu}$. In this case, the taus will be produced by the mediation of
the stau. This corresponds to the tau-dominated scenario
in the CMS analysis~\cite{CMS_chi}.
For a comparison, we pick up one MSSM sample point, calculated with
\textsc{softsusy 3.3.5}~\cite{Allanach:2001kg}, whose mass spectrum
is the similar as our NMSSM benchmark points except $m_{\wt
  \tau} \approx (m_{\wt \chi_1^0} + m_{\wt \chi_2^0}) / 2$,
and then produce the detector-level events by \textsc{Herwig++} and
\textsc{Delphes}. Since the mediating stau is a scalar state,
the angular separation among the final-state leptons is not expected to be
confined in the small $\Delta R_{\ell^+\ell^{\prime-}}$ region
aside from the error due to the mis-paired leptons.
Fig.~\ref{fig:delta_r} demonstrates that the condition
with the collinear leptons is not useful in the typical MSSM point.

Although the $m_{\ell^+\ell^-}^{\rm OSSF}$ cut removes the background
processes associated with $Z$ bosons at least partially, $Z \to
\tau^+\tau^- \to \ell^+\ell^{\prime -} + \MET$ events can be further
reduced by imposing a cut on dilepton invariant masses for all
possible combinations of the isolated leptons ($e,\,\mu$ and $\tau$-jet),
including opposite-sign different-flavor (OSDF) lepton pairs.
\begin{figure}[t!]
  \begin{center}
    \includegraphics[width=0.48\textwidth]{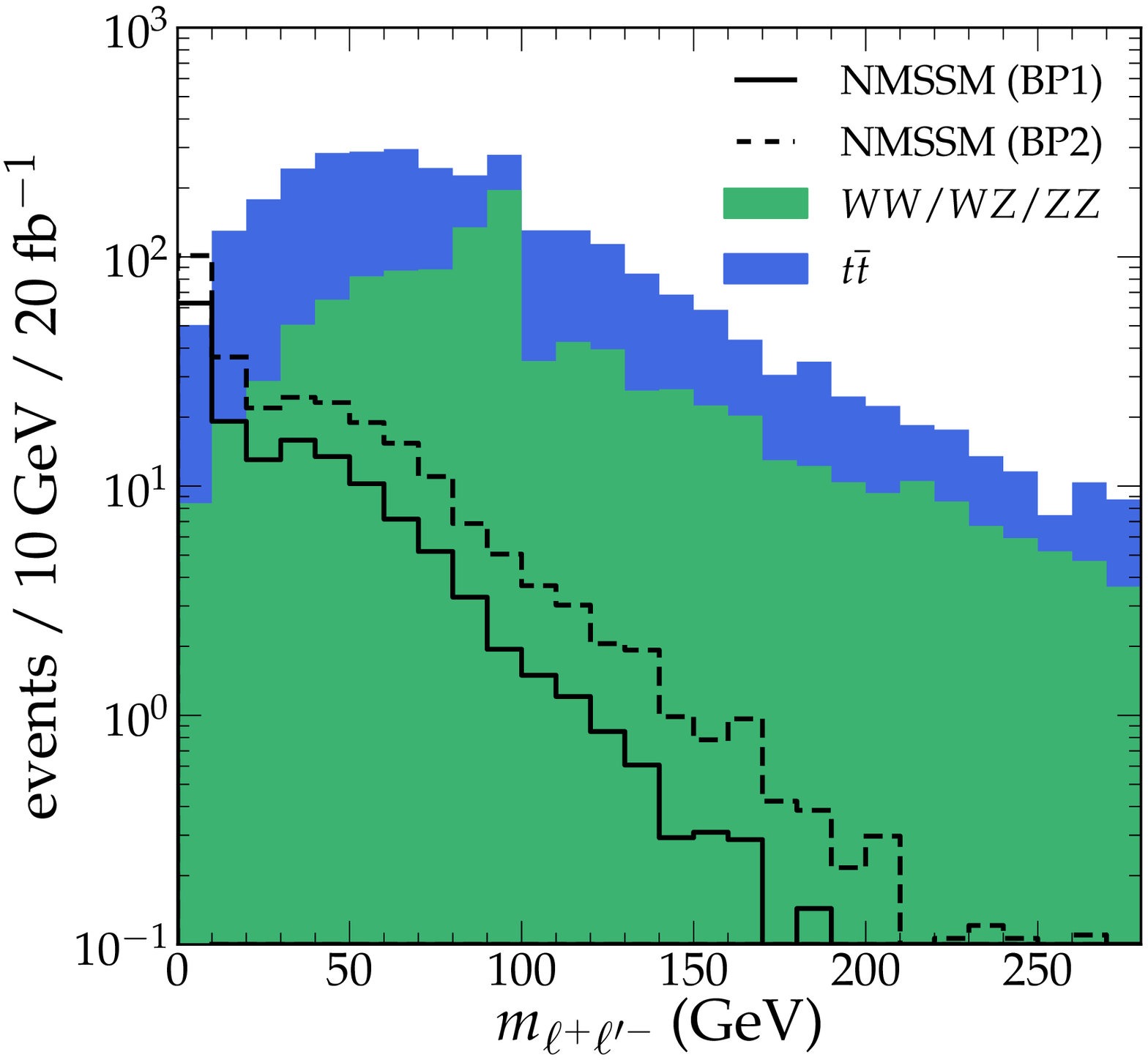}
    \includegraphics[width=0.48\textwidth]{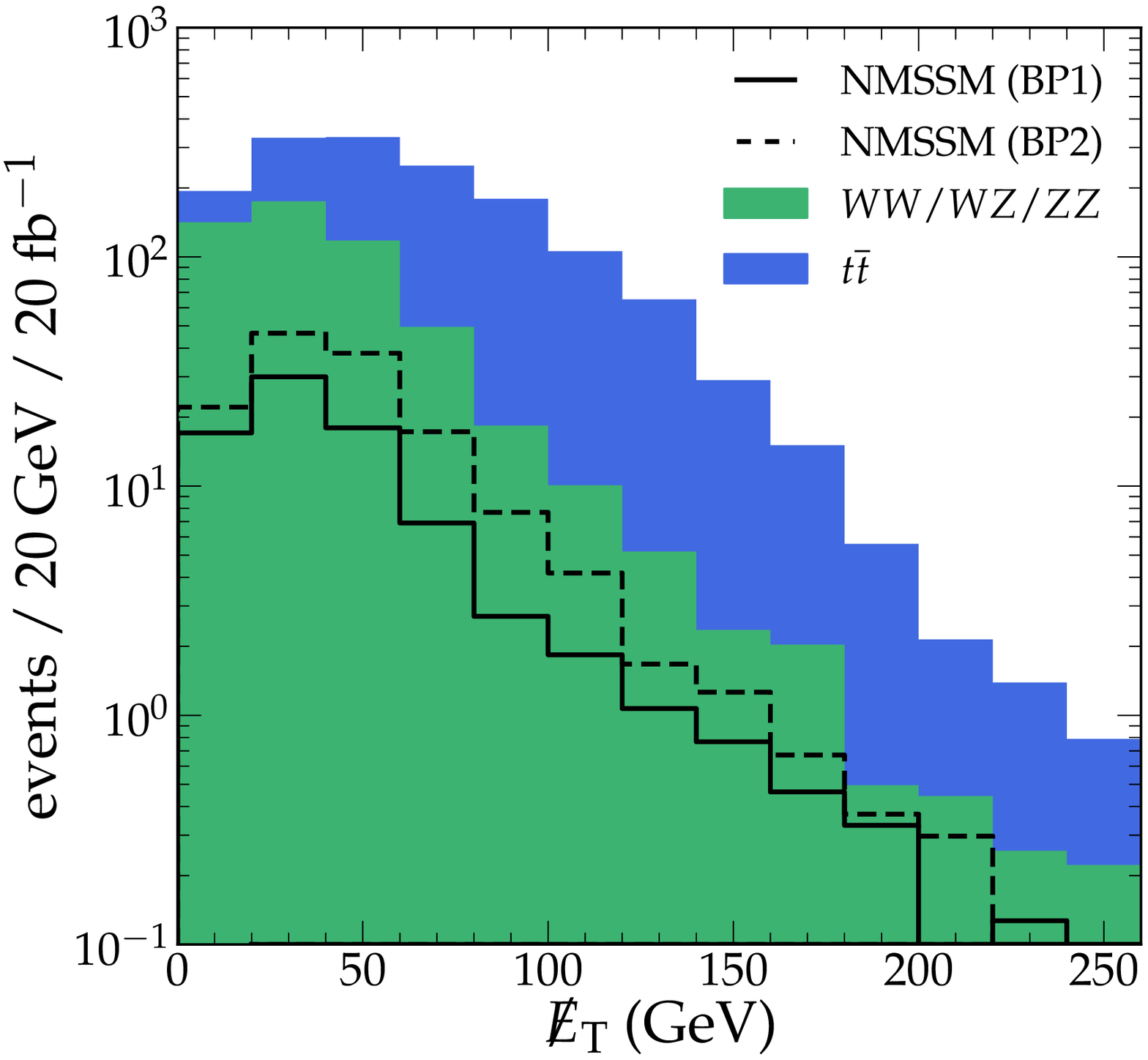}
  \end{center}
  \caption{Distributions of (left panel) dilepton invariant mass for all possible
    combinations of opposite-sign isolated leptons and (right panel)
    the missing energy. Background distributions are normalized to match the
    signal distributions. The basic event selection cuts are applied
    for both signals and backgrounds.}
  \label{fig:m_ll}
\end{figure}
The left panel of Fig.~\ref{fig:m_ll} shows a peak structure around
the $Z$ boson mass in the background distribution, while the SUSY
signals are populated largely in the region of the small mass
value. This observation encourages us to use selection cuts as follows.
\begin{itemize}
  \item If any $m_{\ell^+\ell^{\prime -}}$ for $\ell,\, \ell^\prime =
    e,\, \mu,\, \tau_h$ satisfies $|m_{\ell^+\ell^{\prime -}} - m_Z| <
    20$ GeV, the event is discarded.
  \item $\min_{\forall\{\ell^+\ell^{\prime -}\}} \left\{
    m_{\ell^+\ell^{\prime -}} \right\} < 10$ GeV.
\end{itemize}
The latter cut is set to ensure that at least one pair of leptons originates 
from the light pseudoscalar.
Since similar cuts on $m_{\ell^+\ell^-}^{\rm OSSF}$ have already been
imposed in the basic selection of events,
the cut conditions above are practically applied to the OSDF
lepton pairs.

As mentioned above, the recent searches for the SUSY signature at the
LHC have employed kinematic variables like the missing energy
$\MET$ and the transverse mass $M_{\rm T}$ as well as the dilepton
invariant mass. Strong cuts on these variables can be validated if a
heavy LSP pair is the main source of the missing energy and the mass
gap among sparticles are large enough so that the visible
leptons are very energetic.
However, in our benchmark scenarios, the
missing energy can be quite small due to the cancellation between the LSP and
neutrinos. This is one of distinguishing features of the scenario since the
neutrinos from the tau decay are nearly collinear and the sum of the
neutrino momenta would cancel partially the LSP momenta in the rest
frame of the heavier neutralino as in
Fig.~\ref{fig:kinematic_conf}. This can be checked by seeing the
distributions of the MC events shown in the right panel of
Fig.~\ref{fig:m_ll}.
Still, the $\MET$ cut should be applied to suppress the backgrounds
containing little missing energies like in the QCD multi-jet processes
faking leptons. We here impose rather a mild cut on the missing
energy, $\MET > 30$ GeV.
Moreover, many of the isolated $e,\,\mu$ in the SUSY signal events are
from the tau, which is already quite collinear to the parent light
pseudoscalar, the visible lepton would be soft as discussed
above.\footnote{
  At least one hard isolated lepton can be produced in the leptonic
  decay process of the lighter chargino, i.e., $\wt \chi_1^\pm \to \ell^\pm
  \nu_\ell \wt \chi_1^0$.
}
This situation with the small missing energy and the soft leptons makes
the transverse mass variable less efficient for suppressing the
backgrounds without sacrificing the SUSY signal events.
\begin{figure}[t!]
  \begin{center}
    \includegraphics[width=0.48\textwidth]{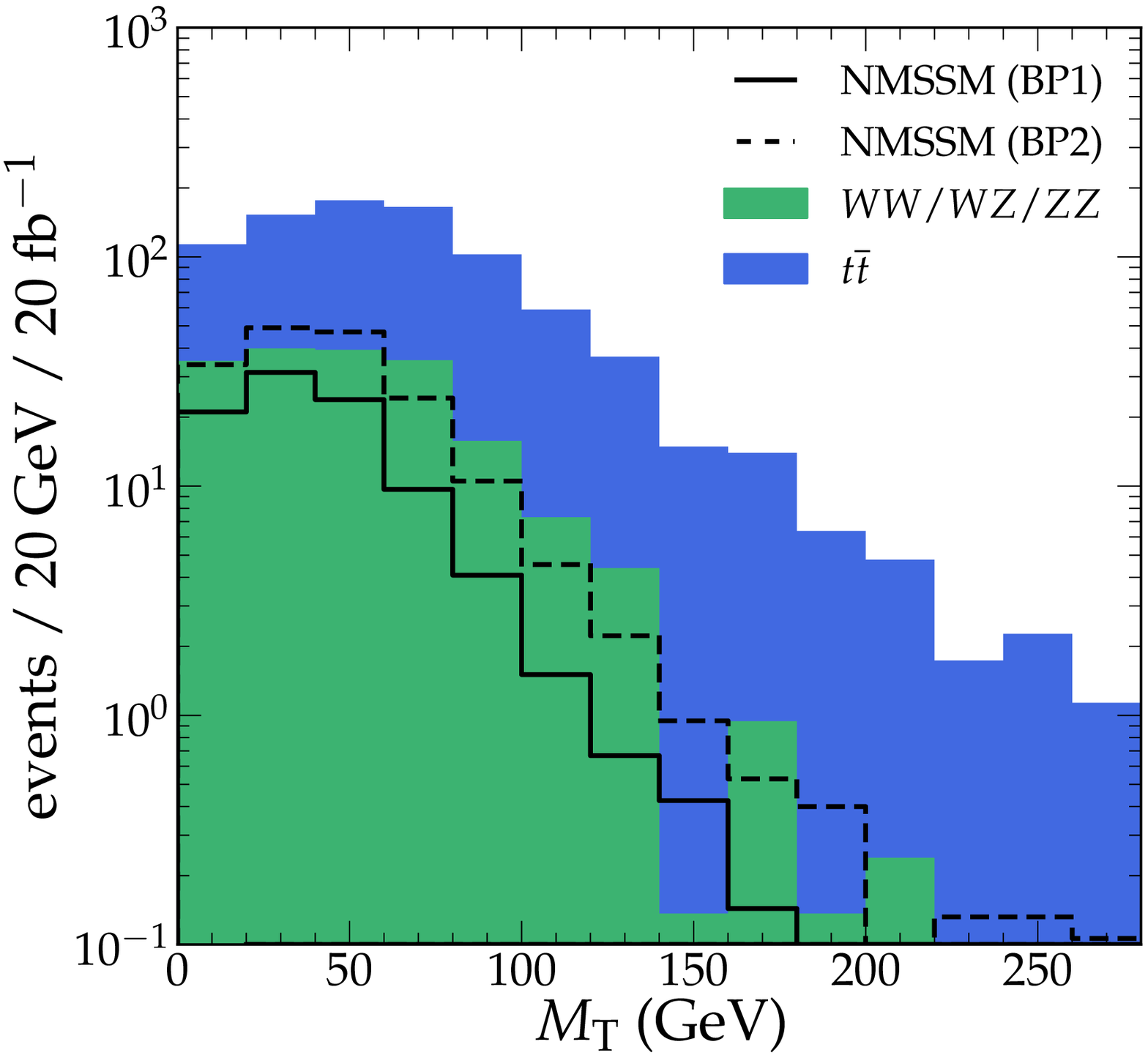}
    \includegraphics[width=0.48\textwidth]{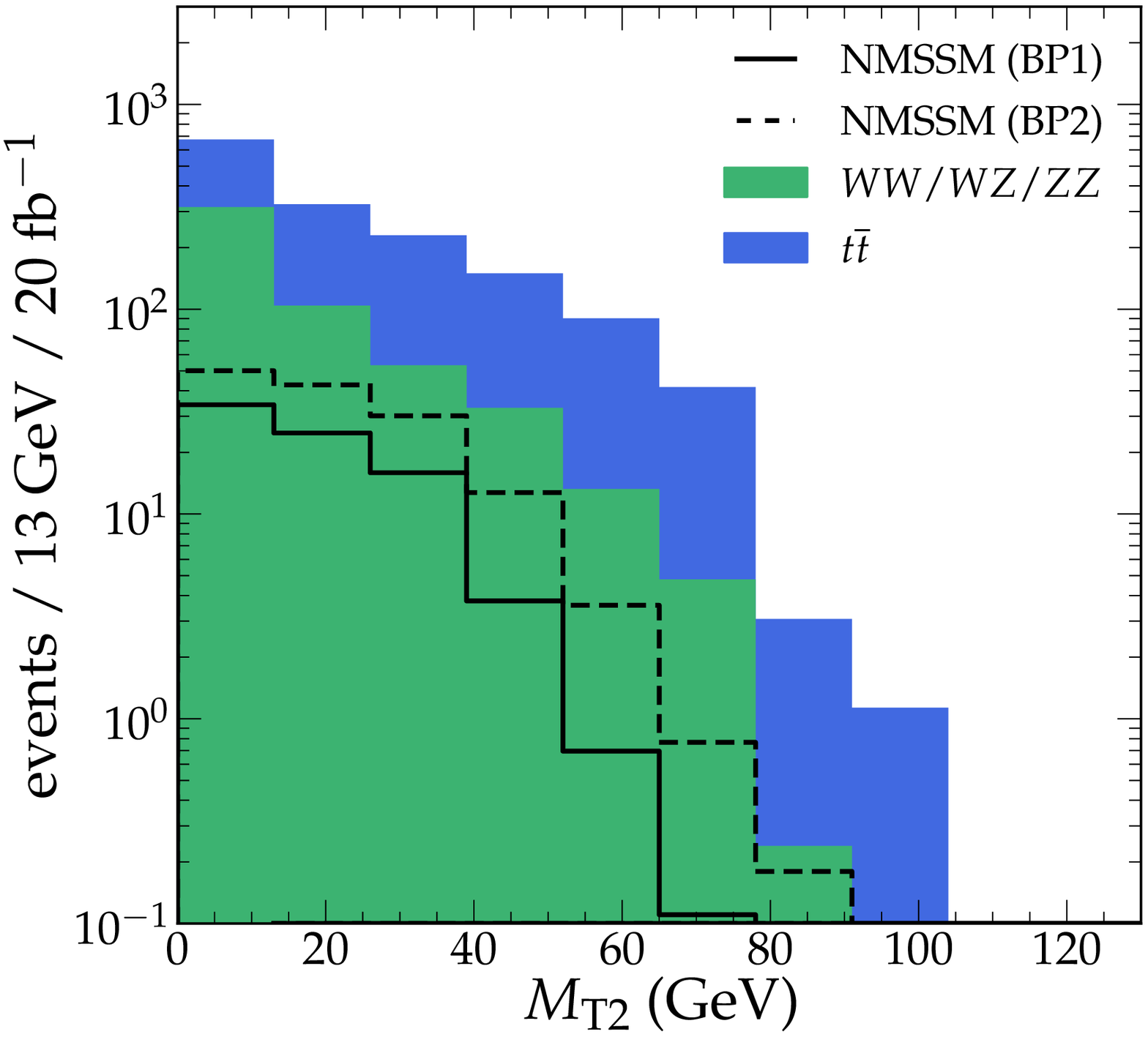}
  \end{center}
  \caption{Distributions of (left panel) transverse mass for one
    lepton $+$ missing energy and (right panel) and the $M_{\rm T2}$
    for an opposite-sign lepton pair and missing energy. Background
    distributions are normalized to match the signal
    distributions. The basic event selection cuts are applied
    for both signals and backgrounds.}
  \label{fig:m_T}
\end{figure}

To see the effect, we compute the transverse mass of one lepton, which
is not paired with the other lepton to obtain the OSSF dilepton
invariant mass, and the missing energy. See the left panel of
Fig.~\ref{fig:m_T}.
Since the signals are populated largely in the smaller region, we
apply an upper cut instead of a lower cut on the transverse mass,
$M_{\rm T} < 60$ GeV.
Since the signal contains multiple leptons, we further attempt to use the
other collider variable $M_{\rm T2}$, which is a generalized transverse
mass variable applicable in the case when there are two invisible
particles in the event, for the system of an opposite-sign lepton pair
and the missing energy~\cite{Lester:1999tx,Barr:2003rg}.
Although it does not have any particular correlation with the
sparticle masses in the signal processes, the dileptonic $t\bar{t}$
and $WW$ backgrounds exhibit the edges around $m_W$ as can be seen in
the right panel of Fig.~\ref{fig:m_T}.
By definition, the input trial mass for the invisible particle is
necessary to calculate the $M_{\rm T2}$.
We set the invisible particle to be massless as it is the correct
choice for the backgrounds where the neutrino is the main source of
the missing energy.
Among several ways of pairing the leptons, we calculate
the $M_{\rm T2}$ of all possible opposite-sign lepton pairs and choose
the smallest value among them in the event. Then, an upper
cut as $M_{\rm T2} < 35$ GeV is imposed similarly as $M_{\rm T}$ since
the $M_{\rm T2}$ values of the signal events are small as can be seen
in the right panel of Fig.~\ref{fig:m_T}.
In the CMS analysis, the $M_{\rm CT}$, so-called contransverse mass
defined in Ref.~\cite{Tovey:2008ui}, has been also used. However, it
is claimed that the $M_{\rm CT}$ is equivalent to the $M_{\rm T2}$ in
the case when the visible and invisible particles are
massless~\cite{Lester:2011nj,Lally:2012uj}. Therefore, we do not apply the cut on
the $M_{\rm CT}$ in our analysis.

\begin{table}[t!]
  \begin{center}{\small
    \begin{tabular}{|l|*{2}{c}|*{6}{c}|}
      \hline&&&&&&&&\\[-2mm]
      Selection cuts & BP1 & BP2 &
      Diboson & Triboson & $t\bar{t}$ & $t\bar{t}W/Z$ & DY & $Z+{\rm jet}$
      \\[2mm]
      \hline&&&&&&&&\\[-2mm]
      Basic cuts &
      79.5 & 140.2 &
      523.6 & 2.1 & 991.4 & 3.0 & 4966.9 & 885.6
      \\[2mm]
      $\MET > 30$ GeV &
      47.2 & ~\,96.0 &
      302.2 & 1.8 & 870.7 & 2.7 & ~\,292.7 & 123.5
      \\[2mm]
      $\min\Delta R_{\ell^+\ell^{\prime -}} < 0.45$ &
      45.0 & ~\,86.8 &
      ~~~\,9.5 & 0.2 & ~\,86.7 & 0.2 & ~~~16.7 & ~~\,7.7
      \\[2mm]
      $m_{\ell^+\ell^{\prime -}}$ cuts &
      30.9 & ~\,55.1 &
      ~~~\,3.1 & ~\,0.04 & ~\,25.5 & ~\,0.03 & ~~~~\,0.0 & ~~\,0.0
      \\[2mm]
      $M_{\rm T}<$ 60 GeV &
      27.9 & ~\,47.2 &
      ~~~\,2.2 & ~\,0.02 & ~\,17.0 & ~\,0.03 & -- & --
      \\[2mm]
      $M_{\rm T2}<$ 35 GeV &
      24.6 & ~\,37.3 &
      ~~~\,1.5 & ~\,0.01 & ~~\,6.2 & ~\,0.02 & -- & --
      \\[2mm]
      Jet-veto &
      16.4 & ~\,25.7 &
      ~~~\,1.4 & ~\,0.01 & ~~\,1.7 & -- & -- & --
      \\[2mm]
      \hline
    \end{tabular}}
  \end{center}
  \caption{Number of events passed the event selection cuts defined in the
    text at the 20 fb$^{-1}$ integrated luminosity.}
  \label{table:cut}
\end{table}

In addition, a jet-veto cut which rejects events containing 
high-$p_{\rm T}$ jets can be applied.
The selection cuts defined in this
section seem to be already good enough for suppressing the leading-order
backgrounds considered in this study. 
However, since we have used the leading-order MC generators to
simulate the SM backgrounds as well as the SUSY signals, a correct
modeling of the higher-order processes like $t\bar{t} +$ jets or $WW
+$ jets could affect the analysis in the real situation. The effect
would not be significant for the SUSY signal events considered
here since they do not have the source of jets in the matrix element
level, except the initial state radiation. In such cases, the
jet-veto cut, which rejects events containing high-$p_{\rm T}$ jets,
would be useful to reduce the multi-jet backgrounds. In order to
provide reference values for the dedicated experimental searches,
we further see the effect of the jet-veto cut with the threshold
$p_{\rm T}$ value of 40 GeV on the signals of neutralinos and chargino
signals on top of the other selection cuts.

To estimate the cut efficiency and the signal significance, we show the
number of events that passed the cumulative event selection cuts discussed
up to now in Table~\ref{table:cut} for both BP1 and BP2. When all the
cuts, including jet-veto condition, are applied, the signal significance is well
above the discovery criterion for both benchmark scenarios.
Therefore, we anticipate
that the SUSY signal of this kind of scenarios can be discovered even in the present 8 TeV LHC
data by tuning the lepton isolation parameters optimized for the
non-standard lepton signals.
%

Finally, we show the importance of a small $\Delta R$ value
in Table~\ref{table:deltacut}. An increase
in $\Delta R$ drastically reduces the number of signal events, evidencing 
the need of choosing a stringent criterion for lepton isolation in order to 
study this scenario. Notice that this implies modification of the search strategies 
usually performed in ATLAS or CMS 
with a dedicated criterion for lepton isolation, optimised for
the decay processes involving light pseudoscalar.


\begin{table}[t!]
  \begin{center}
    \begin{tabular}{| l | *{6}{c} |}
      \hline&&&&&&\\[-2mm]
      \multirow{2}*{Selection cuts} &
      \multicolumn{2}{c}{$\Delta R = 0.1$} &
      \multicolumn{2}{c}{$\Delta R = 0.2$} &
      \multicolumn{2}{c|}{$\Delta R = 0.3$} 
      \\[2mm]
      & BP1 & BP2 & BP1 & BP2 & BP1 & BP2 
      \\[2mm]
      \hline&&&&&&\\[-2mm]
      Basic cuts &
      79.5 & 140.2 & 36.3 & 72.5 & 17.3 & 38.8
      \\[2mm]
      $\MET > 30$ GeV &
      47.2 & ~\,96.0 & 21.5 & 48.0 & 10.2 & 25.8
      \\[2mm]
      $\min\Delta R_{\ell^+\ell^{\prime -}} < 0.45$ &
      45.0 & ~\,86.8 & 19.1 & 41.0 & ~\,8.5 & 19.2
      \\[2mm]
      $m_{\ell^+\ell^{\prime -}}$ cuts &
      30.9 & ~\,55.1 & 13.0 & 24.7 & ~\,5.9 & 11.2
      \\[2mm]
      $M_{\rm T}<$ 60 GeV &
      27.9 & ~\,47.2 & 12.2 & 22.0 & ~\,5.5 & ~\,9.7
      \\[2mm]
      $M_{\rm T2}<$ 35 GeV &
      24.6 & ~\,37.3 & 10.0 & 16.4 & ~\,4.4 & ~\,6.9
      \\[2mm]
      Jet-veto &
      16.4 & ~\,25.7 & ~\,6.4 & 11.4 & ~\,2.9 & ~\,5.0
      \\[2mm]
      \hline
    \end{tabular}
  \end{center}
  \caption{Variation of the number of events that pass 
the selection cuts defined in the
    text at the 20 fb$^{-1}$ integrated luminosity, for various choices of the isolation criterion, $\Delta R$.}
  \label{table:deltacut}
\end{table}

\section{Conclusions}
\label{sec:conclusions}

We have studied potential LHC signatures induced by the presence of
very light pseudoscalar Higgs boson (in the mass range
$2m_\tau<m_{\a1}<2m_b$) in neutralino decays in a scenario 
with two light scalar Higgses within the context of 
the NMSSM. More specifically, we have considered regions of the NMSSM parameter space which feature 
a SM-like Higgs boson in the mass range $123-127$~GeV together with
another lighter one, $\h1$, which is mostly singlet.
For the range of masses considered, the pseudoscalar predominantly decays into a pair of 
taus, $\a1\to \tau^+\tau^-$, leading to an abundance of leptons in the final state.
The resulting LHC phenomenology features multi-lepton signals with missing transverse 
energy in the decay chains which originates from neutralino/chargino pair production,
$\neut_{2,3}\cha_1\to \ell^+\ell^-\ell^\pm+\MET$, 
$\neut_{3}\cha_1\to 2\ell^+2\ell^-\ell^\pm+\MET$, and
$\neut_i\neut_j\to n(\ell^+\ell^-)+\MET$, with $n=2,\,3,\,4$ for $i,j=2,\,3$.

We have performed a scan in the NMSSM parameter space searching for these 
conditions and imposing all the recent experimental constraints on the Higgs 
sector, sparticle masses and low-energy observables. We have further 
assumed that the neutralino is a component of the dark matter and imposed 
the observed upper bound on its relic abundance and on its elastic scattering 
cross section off quarks.
The viable points in the parameter space feature small values of $\kappa$ and $A_\kappa$ 
and as a consequence, light singlet-like $\h1$ and $\a1$.
On top of this, we have also demanded a sizable BR$(\neut_{2,\,3}\to
\neut_1 \a1)$, which favours small values of the $\mu$ parameter since
this leads to Higgsino-like neutralinos. We have distinguished between
two possible scenarios, depending on whether $m_{\h1}<m_{\h2}/2$ or
$m_{\h1}>m_{\h2}/2$, and selected two representative benchmark points.

We have then carried out a reconstruction of the signal for the
selected benchmark points.
The useful cuts with which the signal can be separated from the
background have been determined. After imposing a set
of relevant cuts together with non-standard lepton lepton
separation, the resulting signal to background ratio is
statistically significant at the LHC with 8 TeV center-of-mass energy
and $20~\fbi$ of integrated luminosity. This study suggests that the
analysis of inclusive multilepton searches with missing transverse
energy using the full 8 TeV LHC data and the dedicated selection cuts
can be used to explore corners of the NMSSM parameter space with multiple light
Higgses.

\vspace{-0.2cm}
\subsection*{Acknowledgements}
\vspace{-0.3cm}
\noindent
DGC is supported by the Ram\'on y Cajal 
program of the Spanish MICINN.
CBP is supported by the CERN-Korea fellowship through the National
Research Foundation of Korea.
MP is supported by a MultiDark Scholarship.
DGC, PG and MP thank the support 
of the Consolider-Ingenio 2010 programme under grant 
MULTIDARK CSD2009-00064, the Spanish MICINN under Grants No. FPA2009-08958 
and FPA2012-34694, the Spanish MINECO ``Centro de excelencia Severo Ochoa Program" 
under grant No. SEV-2012-0249, the Community of Madrid under Grant No. HEPHACOS S2009/ESP-1473, 
and the European Union under the Marie Curie-ITN Program No. 
PITN-GA-2009-237920.


\end{document}